\documentclass[12pt]{article}

\usepackage{a4wide,amsfonts,dsfont}
\usepackage{enumerate}
\usepackage[mathscr]{eucal}
\usepackage{graphicx}

\newtheorem{thm}{Theorem}[section]
\newtheorem{lem}{Lemma}[section]

\newtheorem{definition}[thm]{Definition}

\newtheorem{remark}[thm]{Remark}

\newcommand{\R}{\mathbb{R}}

\newcommand{\eje}{\langle\vec{x}\rangle}
\newcommand{\immersion}{\phi:M\longrightarrow\mathbb{L}^3}
\newcommand{\Ltres}{\mathbb{L}^3}

\newenvironment{pf}{\par \noindent\textit{Proof:}\ }{\hfill $\Box$}

\newcommand{\address}[1]{\par \noindent\textsc{#1}\par }
\newcommand{\subjclass}[1]{\noindent \textit{MSC 2000 Classification:}\ #1\\ }
\newcommand{\keywords}[1]{\noindent \textit{Keywords:}\ #1 \\ }

\newcommand{\traza}[1]{\mathrm{trace}(#1)}

\begin{document}

\title{\textbf{Rotational Surfaces in $\mathbb{L}^3$  and Solutions of the Nonlinear Sigma Model}}

\author{Manuel Barros, Magdalena Caballero and Miguel Ortega\thanks{M. Barros' email: mbarros@ugr.es, M. Caballero's  email: mccmm@ugr.es, M. Ortega's email:miortega@ugr.es. This work was partially supported by MEC Grant  MTM2007-60731  with FEDER funds and the Junta de Andaluc\'{\i}a Grant PO6-FQM-01951.}}

\date{}

\maketitle

\address{Departamento de Geometr\'{\i}a y Topolog\'{\i}a, Universidad de Granada, 18071 Granada, Spain}

\begin{abstract}
\noindent The Gauss map of non-degenerate surfaces in the three-dimensional Minkowski space are viewed as dynamical fields of the two-dimensional $O(2,1)$ Nonlinear Sigma Model. In this setting, the moduli space of solutions with rotational symmetry is completely determined. Essentially, the solutions are warped products of orbits of the 1-dimensional groups of isometries and elastic curves in either a de Sitter plane, a hyperbolic plane or an anti de Sitter plane. The main tools are the equivalence of the two-dimensional $O(2,1)$ Nonlinear Sigma Model and the Willmore problem, and the description of the surfaces with rotational symmetry. A complete classification of such surfaces is obtained in this paper. Indeed, a huge new family of Lorentzian rotational surfaces with a space-like axis is presented. The description of this new class of surfaces is based on a technique of surgery and a gluing process, which is illustrated by an algorithm.

\end{abstract}


\section{Introduction}

Nonlinear sigma models are field theories whose
elementary fields, or dynamical variables, are maps, $\phi$, from
a space, $M$, the source space, to an auxiliary space,
$\mathbf{E}$, the target space, endowed with a non-degenerate metric. The Lagrangian governing the
dynamics of the model measures the total energy of those maps. The
classical solutions of the model, i.~e., the solutions of the
corresponding field equations, constitute the space of field
configurations. The
dimension of the source space is called the dimension of the model. The isometry group, $\mathbf{A}$, of the target space is the
symmetry of the model. In particular, when $M$ is compact and Riemannian, each solution has finite energy. In this sense, we call them solitons.

Two-dimensional nonlinear sigma models, in particular those with
symmetry $O(3)$ and $O(2,1)$, are ubiquitous in Physics (see for
example \cite{Cavalcante,Tsurumaru} and references therein); with
applications going from Condensed-matter Physics (see
\cite{Belavin-Polyakov,Laughlin,Mieck} and references therein) to
High-energy Physics (see \cite{Ceciliauno,Ceciliados,Howe,Otsu}
and references therein) and, of course, Quantum Field Theory (see
\cite{Mishchenko,Schutzhold} and references therein). 
In particular, those with Minkowski signature metric on the target space are applied to 
Gauge Theories (see \cite{Ceciliauno,Tseytlin1}), Quantum Gravity (see \cite{Tseytlin2}), String Theories (see \cite{Bredthauer,Tseytlin2}), Quantum Mechanics (see \cite{Davis2}) and General Relativity, in particular Einstein and Ernst equations (see \cite{Chelnokov,Gruszczak}). They are
specially important in string theories where the model description
is applicable. This kind of universality is strongly related to
the fact that these sigma models, and the equations governing their
dynamics, have a deep underlying geometric meaning. This provides
a powerful reason to explain the great interest of these models in
Applied Mathematics and in Differential Geometry, even
without mentioning any physical terminology, simply as a kind of
constrained Willmore problem (see, for example, \cite{Anzellotti1,Anzellotti2,Bracken,Capovilla} and references therein). In this framework, it
seems natural to identify the dynamical variables of the 
two-dimensional $O(3)$ Nonlinear Sigma Model 
with the Gauss maps of surfaces in the three-dimensional Euclidean space. This
approach has been successfully used to obtain certain moduli
spaces of solutions: with constant mean curvature,
\cite{Belavin-Polyakov,DoCarmo-Dajzer,Ody,Purkait-Ray}, those
admitting a rotational symmetry, \cite{Barros1}, and those
foliated by Villarceau circles, \cite{Barros-Caballero-Ortega}.

The study of moduli spaces of solutions (field configurations) of the two-dimensional $O(2,1)$ Nonlinear Sigma Model constitutes an ambitious program. We will develop it along a series of articles, starting with this one. Beforehand, it will be useful to remark the following general points related to this model:

\begin{itemize}
 \item The Gauss map of any nondegenerate surface in the three-dimensional Lorentz-Minkowski space, $\Ltres$, is automatically an elementary field of this model. Therefore, the geometrical approach identifies the space of dynamical variables with that of Gauss maps of nondegenerate surfaces in $\Ltres$.

\item On the other hand, the underlying variational problem of this model turns out to be equivalent to the Willmore variational problem (see Theorem \ref{equivalencia}). This has important consequences:
\begin{enumerate}
 \item The field configurations of this model are nothing but the Willmore surfaces in $\Ltres$.
\item The model is invariant under conformal changes of the metric of $\Ltres$.
\item Since the Willmore functional is essentially the Polyakov action, the model can be regarded as a bosonic string theory in $\Ltres$ that is governed by the Willmore-Polyakov action. In this sense, the solutions of the model provide the string world sheet configurations (see \cite{Pol1,Pol2}).
\end{enumerate}

\end{itemize}

Now, the first step in the above program, which constitutes the
main aim of this paper, is stated as follows: 
\begin{quote}
\textbf{To determine the moduli space of solutions of the
two-dimensional $O(2,1)$ Nonlinear Sigma Model that admit a
rotational symmetry. Equivalently, classify, up to congruences,
those rotational surfaces in $\mathbb{L}^3$ that are critical
points of the total energy.}
\end{quote}
This problem is much more difficult and subtle than its
Riemannian partner, \cite{Barros1}, and it will be treated
according to the causal character of the symmetry axis.
Indeed, in Section \ref{Auno-solutiones}, we have studied and completely solved
the case where the symmetry axis is time-like, that is, surfaces
invariant under a one parameter group, $\mathbf{A}_1$, of
elliptic motions. This can be summarized as follows:
\begin{enumerate}
    \item Firstly, we consider the nonlinear sigma model with
    boundary and we determine the admissible boundary conditions.
    \item Next, we obtain the space of surfaces that are
    invariant under rotations with time-like axis.
    \item Then, since the orbits are circles, we use the principle
    of symmetric criticality, \cite{Palais}, and the conformal
    invariance of the model to make a suitable conformal change to obtain that
\end{enumerate}
\begin{quote}
\textbf{The solutions of the two-dimensional $O(2,1)$ Nonlinear
Sigma Model that admit a rotational symmetry with time-like axis are
obtained by rotating clamped free elastic curves (critical points of the total squared curvature) in the anti de Sitter
plane}.
\end{quote}

The major part of the paper is devoted to obtaining rotational
solutions with space-like axis. This case
is the most complicated.

The first important difficulty is to obtain the whole class
of rotational surfaces in $\mathbb{L}^3$ with space-like axis, in
other words, surfaces that are invariant under a one parameter
group, $\mathbf{A}_2$, of hyperbolic motions. This problem, which
has been usually avoided in the literature, perhaps because of its
difficulty, is completely solved in Section \ref{superficiesAdos}.
To understand this problem, assume that the space-like axis
coincides with the $\{x\}$-axis. Then, the planes $y=z$ and $y=-z$
divide $\mathbb{L}^3$ in four open regions, which will be called
fundamental regions. Certainly, for every fundamental region we can get a class of rotational
surfaces, with $\{x\}$-axis, immersed in the region.  These surfaces are well known in the literature (see for
example \cite{Hano-Nomizu}). However, there are rotational surfaces with $\{x\}$-axis in $\mathbb{L}^3$ that
leave a fundamental region to emerge in another fundamental
region. This family includes popular surfaces, such as a saddle
surface and the one-sheet hyperboloid with $\{z\}$-axis. In some
sense, these surfaces can be obtained by gluing two or more surfaces, each of them contained in a fundamental region. Along Section \ref{superficiesAdos}, we use surgery to dissect these surfaces and to understand the gluing mechanism. At the end of it, we obtain a classification theorem (see
Theorem \ref{Lorentzianas-Invariantes}) and a construction
algorithm (see Subsection \ref{algoritmo}).

Once we have obtained the whole space of rotational surfaces with
space-like axis, there are, at least, two different ways to get the
corresponding solutions. On one hand, one can try to carry out a 
symmetry reduction of the action principle. This procedure depends on a kind of symmetric
criticality principle that should be established. However, there
is a second way that consists in a direct variational approach.
Therefore, one needs to obtain the field equations governing the
model. Since the model turns out to be equivalent to a constricted
Willmore model, in Section \ref{variacion-will}, we obtain the
first variation that provides Willmore surfaces in a general
semi-Riemannian background.

In Section \ref{Ados-R-solutiones}, we obtain the whole moduli space of
Riemannian solutions with a rotational symmetry with space-like axis(see
Theorem \ref{Riemannian-solutions}).

\begin{quote}
\textbf{The Riemannian solutions of the two-dimensional $O(2,1)$
Nonlinear Sigma Model that admit a rotational symmetry with space-like axis
are obtained by rotating space-like clamped free elastic curves of
the de Sitter plane}.
\end{quote}

In Section \ref{Ados-L-solutiones}, the whole moduli space of
Lorentzian solutions with a rotational symmetry with space-like axis is
obtained. Firstly, we study those solutions that are contained in a
fundamental region (fundamental solutions).

\begin{quote}
\textbf{On one hand, we get solutions coming from time-like clamped
free elastic curves in the de Sitter plane} (Theorem
\ref{Lorentzian-Fundamental-solutions-1}). \textbf{On the other hand,
we also obtain a second family of Lorentzian solutions, which are
generated by clamped free elastic curves in the hyperbolic plane}
(Theorem \ref{Lorentzian-Fundamental-solutions-2}). 
\end{quote}

Certainly, each
solution in those families is contained in a fundamental region. In
contrast with the Riemannian case, we can find Lorentzian solutions
in all the fundamental regions. 
This fact allows us to study the existence of solutions leaving a
fundamental region and emerging in another one. In other words, we
look for solutions obtained by gluing fundamental solutions. This
problem is completely solved at the end of Section \ref{Ados-L-solutiones}. In
fact, such solutions are surfaces that are connected pieces of either a one-sheet hyperboloid with time-like axis and centered
at any point of the space-like axis, or a Lorentzian plane
orthogonal to the space-like axis (Theorem
\ref{Lorentzian-not-fundamental-solutions}).

Finally, in the last section we consider the case where solutions
admit a one parameter group, $\mathbf{A}_3$, of parabolic
transformations. They are known in the literature as
rotational surfaces with light-like axis, \cite{Hano-Nomizu}. Now,
parabolic rotational surfaces lie in two fundamental regions of
$\mathbb{L}^3$. In contrast with the case of rotational surfaces
with space-like axis, in Section \ref{superficiesAtres} we prove
that we can not find parabolic rotational surfaces that leave a
fundamental region to emerge in the other. At the end of that
section, we obtain the complete classification of
$\mathbf{A}_3$-invariant solutions (see Theorem \ref{G3-solutions}).

\begin{quote}
\textbf{The solutions of the two-dimensional $O(2,1)$ Nonlinear Sigma
Model that admit a rotational symmetry with light-like axis are obtained by
rotating clamped free elastic curves of the anti de Sitter plane}.
\end{quote}

The results of this paper can be summarized in the following
statement.

\begin{quote}
\textbf{The solutions of the two-dimensional $O(2,1)$ Nonlinear
Sigma Model that admit a rotational symmetry are the following surfaces:
\begin{enumerate}
    \item A connected piece (with boundary) of a Lorentzian plane.
    \item A connected piece (with boundary) of a one-sheet hyperboloid with time-like axis.
    \item A surface generated, via rotations, by a clamped free 
    elastic curve according to the following table.\end{enumerate}}
\end{quote}

\vspace{1em}

\begin{center}

\begin{tabular}{|p{5.5em}|l|l|l|p{10em}|}
\hline

\textbf{Symmetry Group} & \textbf{Axis} &\textbf{Orbits}

&\textbf{Character of the surface}  &\textbf{Generating Curve} \\

\hline\hline

$\mathbf{A}_1$ & Time-like & Circles & Riemannian & Space-like
free elastic curve in the  anti de Sitter plane  \\

\hline

$\mathbf{A}_1$ & Time-like & Circles & Lorentzian & Time-like
free elastic curve in the anti de Sitter plane  \\

\hline

$\mathbf{A}_2$ & Space-like & Hyperbolas & Riemannian &
Space-like free elastic curve in the de Sitter plane  \\

\hline

$\mathbf{A}_2$ & Space-like & Hyperbolas & Lorentzian &
Time-like free elastic curve in the de Sitter plane  \\

\hline

$\mathbf{A}_2$ &  Space-like & Hyperbolas & Lorentzian & Free elastic curve in the hyperbolic plane  \\

\hline

$\mathbf{A}_3$ & Light-like & Parabolas & Riemannian & Space-like free elastic curve in the anti de Sitter plane  \\

\hline

$\mathbf{A}_3$ &  Light-like & Parabolas & Lorentzian & Time-like free elastic curve in the anti de Sitter plane  \\

\hline
\end{tabular}
\end{center}

\noindent \textit{Acknowledgement.} The authors would like to thank the referee for their useful comments, which helped us to improve this paper.

\newcommand{\Np}{N_{\phi}}

\section{Preliminaries and Generalities} \label{prel}

Along this paper the geometrical objects are $C^{\infty}$ or
equivalently smooth, though those appearing in the paper could be
supposed to be only as differentiable as needed.

Let $M$ be a surface with (or without) boundary
and $\phi:M\longrightarrow\mathds{L}^3$ an immersion in the
Lorentz-Minkowski three space with flat metric $g=\langle,\rangle$.
If $d\phi_p(T_p M)$ is a non-degenerate plane in $\mathds{L}^3$
for any $p\in M$, then
$\phi:(M,\phi^{*}(g))\longrightarrow\mathds{L}^3$ is said to be a
non-degenerate isometric immersion (or a non degenerate surface).
A non-degenerate surface can be oriented, at least locally, by a
unitary normal vector field, say $\Np$. According to the causal
character, $\langle \Np,\Np\rangle=\varepsilon$, we have two
possibilities:
\begin{description}
    \item[Lorentzian surfaces (also called time-like surfaces).] When $\varepsilon=1$,  $\Np$ is space-like and so
$(M,\phi^{*}(g))$ is a Lorentzian surface. The unitary normal vector
field, $\Np$, can be viewed as a map, the Gauss map,
$\Np:M\longrightarrow\mathds{S}_1^2$, where
$\mathds{S}_1^2=\{\vec{v}\in\mathds{L}^3\,:\,\langle
\vec{v},\vec{v}\rangle=1\}$ is the de Sitter plane.

    \item[Riemannian surfaces (also named space-like surfaces).] When $\varepsilon=-1$,  $\Np$ is time-like and so
$(M,\phi^{*}(g))$ is a Riemannian surface. In this case, the Gauss
map is defined as $\Np:M\longrightarrow\mathds{H}^2$, where
$\mathds{H}^2=\{\vec{v}\in\mathbf{C}^{\uparrow}\subset\mathds{L}^3\,:\,\langle
\vec{v},\vec{v}\rangle=-1\}$ is the hyperbolic plane  and
$\mathbf{C}^{\uparrow}$ denotes the future cone.
\end{description}

Let us denote by $O(2,1)$ the matrix representation of the group of
vectorial isometries of $\mathds{L}^3$,
$\textrm{Iso}(\mathds{L}^3)$, also known as the group of Lorentz
transformations.  Since the group of isometries of both
$\mathds{S}_1^2$ and $\mathds{H}^2$ is $O(2,1)$, the Gauss map of
non-degenerate surfaces can be regarded as elementary fields in the
two-dimensional $O(2,1)$ Nonlinear Sigma Model (sometimes we will
abbreviate it as $O(2,1)$ NSM). The Lagrangian density governing
this field theory is precisely $\|d\Np\|^2$, which can be computed,
via the Gauss equation, in terms of the mean curvature, $H_{\phi}$,
of $(M,\phi)$ and the Gaussian curvature, $K_{\phi}$, of
$(M,\phi^{*}(g))$, i. e.,
\begin{equation}\label{Gauss-Equation}
    \|dN_{\phi}\|^2=4H_{\phi}^2-2\varepsilon K_{\phi}.
\end{equation}

Consider
$\textrm{Iso}^{+\uparrow}(\mathds{L}^3)=\{f\in\textrm{Iso}(\mathds{L}^3)\,:\,
\textrm{det}(f)=1,
f(\mathbf{C}^{\uparrow})=\mathbf{C}^{\uparrow}\}$. Its partner in
$O(2,1)$ is denoted by $(O(2,1))^{+\uparrow}$. It is known that each
Lorentz transformation, $f\in\textrm{Iso}(\mathds{L}^3)$, admits at
least one eigenvector with eigenvalue $\pm 1$. Therefore, given
$\vec{x}\in\mathds{L}^3$, we wish to determine those Lorentz
transformations, $f\in\textrm{Iso}^{+\uparrow}(\mathds{L}^3)$, such
that $f(\vec{x})=\vec{x}$. These vectorial isometries constitute a
subgroup, $\mathbf{A}$, of $\textrm{Iso}^{+\uparrow}(\mathds{L}^3)$,
called the \textit{group of rotations with axis}
$\eje=\textrm{Span}\{\vec{x}\}$. Certainly, $\mathbf{A}$ acts
naturally on the whole $\mathds{L}^3$ producing orbits. However,
these orbits are quite different according to the causal character
of the axis. Next we summarize the corresponding discussion.
\begin{description}
    \item[(1) Time-like axis.] Choose an orthonormal basis, in $\mathds{L}^3$, as
    follows $\mathcal{B}=\{\vec{x},\vec{y},\vec{z}\}$. We work in coordinates with respect to $\mathcal{B}$. Since the
    axis is time-like, then $\{\vec{y},\vec{z}\}$ determine an
    Euclidean plane. In this case, the group
    $\mathbf{A}$ is identified with the following subgroup of $(O(2,1))^{+\uparrow}$,
    \[\mathbf{A}_1 =\{1\}\times O^{+}(2)=\left\{\mu_t=\left(%
\begin{array}{ccc}
  1 & 0 & 0 \\
  0 & \cos{t} & -\sin{t} \\
  0 & \sin{t} & \cos{t} \\
\end{array}%
\right) \, : \, t\in\mathds{R}\right\}.\] Given a point
$p=(a_1,a_2,a_3)\in\mathds{L}^3$, denote by $P$ the Euclidean plane
in $\mathds{L}^3$ passing through $p$ and orthogonal to $\vec{x}$.
The orbit  of $p$ under the action of $\mathbf{A}_1$, $[\,p\,]_1$,
is just the circle in $P$
 through $p$ with center $(a_1,0,0)$, i.~e., $[\,p\,]_1=\{(a_1,y,z)\in\mathds{L}^3:y^2 +z^2=a_{2}^2 + a_{3}^2\}.$
 Certainly, $[\,p\,]_1=\{p\,\}$ if $a_2=a_3=0$. Therefore,
 sometimes we call transformations in $\mathbf{A}_1$ elliptic
 motions or pure rotations.

    \item[(2) Space-like axis.] Choose an orthonormal basis, in $\mathds{L}^3$, as
    follows $\mathcal{B}=\{\vec{x},\vec{y},\vec{z}\}$. We work in coordinates with respect to this basis.
    Since the
    axis is space-like, then $\{\vec{y},\vec{z}\}$ determine a
    Lorentzian plane. In this case, the group
    $\mathbf{A}$ is identified with the following subgroup of $O_1^{+\uparrow}(3)$,
\begin{equation}\label{grupoA2}
\mathbf{A}_2=\{1\}\times O_1^{+\uparrow}(2)=\left\{\xi_t=\left(%
\begin{array}{ccc}
  1 & 0 & 0 \\
  0 & \cosh{t} & \sinh{t} \\
  0 & \sinh{t} & \cosh{t} \\
\end{array}%
\right) \, : \, t\in\mathds{R}\right\}.
\end{equation}
Given a point $p=(a_1,a_2,a_3)\in\mathds{L}^3$, denote by $P$ the
Lorentzian plane in $\mathds{L}^3$ passing through $p$ and
orthogonal to $\vec{x}$. It is clear that the orbit of $p$,
$[\,p\,]_2$, is contained in $P$. If $a_2=a_3=0$, then
$[\,p\,]_2=\{p\,\}$. Otherwise, we must distinguish two cases:
\begin{itemize}
    \item If $a_2^2=a_3^2$, then $[\,p\,]_2$ is the open half straight line
    starting at $(a_1,0,0)$ and passing through $p$, i.~e.,
    $[\,p\,]_2=\{(a_1,\lambda\, a_2,\lambda\, a_3)\,:\,0<\lambda\}$.
    \item If $a_2^2\neq a_3^2$, then $[\,p\,]_2$ is the
    branch of hyperbola in $P$ centered at $(a_1,0,0)$ and
    passing through $p$, i.~e., $[\,p\,]_2$ is the connected component of
    $\{(a_1,y,z)\in\mathds{L}^3\,:\,y^2-z^2=a_2^2-a_3^2\}$
    that contains $p$. Transformations in $\mathbf{A}_2$ will be
    usually called hyperbolic motions or hyperbolic rotations.
\end{itemize}

    \item[(3) Light-like axis.] If $\vec{x}$ is null, then we
    consider a basis $\mathcal{B}=\{\vec{x},\vec{y},\vec{z}\}$ of
    $\mathds{L}^3$ such that: (1) $\vec{y}$ is a light-like vector
    with $\langle \vec{x},\vec{y}\rangle=-1$, and (2) $\vec{z}$ is
    a unitary space-like vector orthogonal to the plane
    $\textrm{Span}\{\vec{x},\vec{y}\}$. We work in coordinates with respect to $\mathcal{B}$. It can be checked that the group
    $\mathbf{A}$ is identified with the following subgroup of $O_1^{+\uparrow}(3)$,
    \[\mathbf{A}_3=\left\{\varsigma_t=\left(%
\begin{array}{ccc}
  1 & \frac{1}{2}t^2 & t \\
  0 & 1 & 0 \\
  0 & t & 1 \\
\end{array}%
\right) \, : \, t\in\mathds{R}\right\}.\]
 In this setting, to analyze the orbits, we must consider again a
 couple of cases different to that where $a_2=a_3=0$, in which
 $[\,p\,]_3=\{p\,\}$, for
 $p=(a_1,a_2,a_3)\in\mathds{L}^3$.
 \begin{itemize}
    \item If $a_2=0$, then the orbit $[\,p\,]_3$
    is a straight line, namely $[\,p\,]_3=\{(t,0,a_3)\,:\,
    t\in\mathds{R}\}$.
    \item If $a_2\neq 0$, then the orbit  $[\,p\,]_3$ is a parabola in the plane
    $Q=\{(x,a_2,z)\in\mathds{L}^3\,:\,x,z\in\mathds{R}\}$. In
    fact, $[\,p\,]_3=\{(x,a_2,z)\in Q\,:\,
    x=\frac{1}{2a_2}(z-a_3)^2+\frac{a_3}{a_2}(z-a_3)+a_1\}$. Thus,
    $\mathbf{A}_3$ will be called the group of parabolic rotations
    or parabolic motions.
 \end{itemize}
\end{description}

To end this section, we define elastic curves, known also as elasticae. We consider a Riemannian or Lorentzian oriented 2-manifold $(M,h)$, with its Levi-Civita connection $\nabla$. We only work with Frenet curves, i.~e., regular curves $\alpha:I\subset\mathds{R}\rightarrow M$ such that the following \textit{Frenet} equations are well-defined,
$$ \nabla_{T} T = \epsilon_2 \kappa  N, \quad \nabla_{T}N = -\epsilon_1 \kappa T,
$$ 
where $\{T=\alpha'/\|\alpha'\|,N\}$ is a positive orthonormal frame along $\alpha$,  $\epsilon_1=h(T,T)=\pm 1$, $\epsilon_2=h(N,N)=\pm 1$ 
and $\kappa$ is a smooth function, usually called the (geodesic) curvature of $\alpha$. We recall that geodesics are curves such that $\kappa$ vanishes identically.

Next, given two points $p_1$, $p_2\in M$ and two tangent vectors $v_i\in T_{p_i}M$, $i=1,2$, we define a space of  \textit{clamped} curves 
$$ \Gamma =\{\alpha:[a_1,a_2]\rightarrow M : 
\alpha(a_i)= p_i, \, \alpha'(a_i)=v_i, \, i=1,2\}.
$$
We also admit the case $p_1=p_2$ and $v_1=v_2$, and then, we call $\Gamma$ a space of \textit{closed} curves. Let us consider the \textit{total squared action} on $\Gamma$, 
$$ \mathfrak{E} : \Gamma \rightarrow \mathds{R}, \quad
\mathfrak{E}(\alpha) =  \int_{\alpha} (\kappa^2+\lambda),
$$
where $\lambda\in\mathds{R}$ is a Lagrange multiplier. Thus, an \textit{elastica} (or an elastic curve) is a critical point of $\mathfrak{E}$. In case $\lambda=0$, we call it \textit{free elastica}.

\section{Gaussian map approach and the Conformal Invariance of the two-dimensional $O(2,1)$ NSM} \label{confo}

The elementary fields in the two-dimensional $O(2,1)$ Nonlinear
Sigma Model are $\mathds{L}^3$-valued unitary vector fields on
surfaces with (or without) boundary. Therefore, they map something
of dimension two (a surface) in either a de Sitter plane,
$\mathds{S}_1^2$, or a hyperbolic one, $\mathds{H}^2$. Along this
paper, we will assume that the source space is a surface with
boundary and elementary fields are subject to a natural constraint
along the boundary. However, the free case, i.~e., when the boundary
of the surface is empty, can be regarded as a particular case with
no constraints. Hence, it seems natural to approach the study of
this sigma model, in connection with the differential geometry of
surfaces in $\mathds{L}^3$, by identifying the dynamical variables
of the model with the Gauss map of non-degenerate surfaces in the
Lorentz-Minkowski three-space. To be precise, let us state the
general setting of this approach.

Let $\Gamma=\{\gamma_1,\gamma_2,\dots,\gamma_n\}$ be a finite set of
non-null regular curves in $\mathds{L}^3$ with
$\gamma_i\bigcap\gamma_j=\emptyset$, if $i\neq j$. Let $N_o$ be a
unitary vector field along $\Gamma$  orthogonal to $\Gamma$ and with constant causal character on the whole $\Gamma$, i.~e., 
\begin{eqnarray*}
  \langle \Gamma'(p),N_o(p)\rangle &=& 0, \quad \forall p\in\Gamma, \textrm{ and} \\
  \langle N_o(p),N_o(p)\rangle &=& \varepsilon, \quad \forall
  p\in\Gamma, \textrm{ where } \varepsilon\in\{1,-1\}.
\end{eqnarray*}
Notice that, if $\varepsilon=1$, $\Gamma$ could consist of both
time-like and space-like curves at the same time. Furthermore,
$\Gamma'$ and $N_o$ determine a third vector field along $\Gamma$
given by $N_o=\Gamma'\wedge\nu.$

On the other hand, let $M$ be a connected smooth surface with
boundary, $\partial M=c_1\cup c_2\cup\dots\cup c_n$. We denote by
$\mathbf{I}_{\Gamma}^{\varepsilon}(M,\mathds{L}^3)$ the space of
immersions, $\phi:M\longrightarrow\mathds{L}^3$ with unitary normal
vector field, $N_{\phi}$, satisfying $\langle
N_{\phi},N_{\phi}\rangle=\varepsilon$ and the following boundary
conditions: 
\begin{enumerate}
    \item $\phi(\partial M)=\Gamma$, namely $\phi(c_j)=\gamma_j$,
    $1\leq j\leq n$, and
    \item $d\phi_q(T_q M)$ is orthogonal to $N_o(\phi(q))$, $\forall
    q\in \partial M$; this is equivalent
     to say $\left. N_{\phi}\right\vert_{\Gamma}=N_o$.
\end{enumerate}
\noindent Roughly speaking, if we identify each immersion $\phi\in
\mathbf{I}_{\Gamma}^{\varepsilon}(M,\mathds{L}^3)$ with its graph,
 $\phi(M)$, viewed as a surface with boundary in $\mathds{L}^3$,
 then $\mathbf{I}_{\Gamma}^{\varepsilon}(M,\mathds{L}^3)$ can be
 regarded as the space of immersed surfaces in $\mathds{L}^3$ having the same causal character,  the same
 boundary and the same Gauss map along the common boundary.

In this setting, the action governing the two-dimensional $O(2,1)$
Nonlinear Sigma Model,
$\mathfrak{S}:\mathbf{I}_{\Gamma}^{\varepsilon}(M,\mathds{L}^3)\longrightarrow\mathds{R}$,
can be written as
\begin{equation}\label{Action}
   \mathfrak{S}(\phi)=\int_{M}\,\|dN_{\phi}\|^2\,dA_{\phi},
\end{equation}
where $dA_{\phi}$ denotes the element of area of
$(M,\phi^{*}(g))$. The solutions of the
two-dimensional $O(2,1)$ Nonlinear Sigma Model are just the
critical points of
$\left(\mathbf{I}_{\Gamma}^{\varepsilon}(M,\mathds{L}^3),\mathfrak{S}\right)$.

Next, we define a \textit{non-null polygon} as a connected,
simply-connected,  compact domain $\mathsf{K}\subset M$ with nonempty
interior and with piecewise smooth boundary, $\partial \mathsf{K}$,
made up of a finite number of smooth non-null curves. The concept of
solution can be materialized according to the following

\begin{definition} \label{solution}$\phi\in\mathbf{I}_{\Gamma}^{\varepsilon}(M,\mathds{L}^3)$
is a critical point of
$\left(\mathbf{I}_{\Gamma}^{\varepsilon}(M,\mathds{L}^3);\mathfrak{S}\right)$
if for any non-null polygon $\mathsf{K}\subseteq M$, the restriction
$\phi\vert_{\mathsf{K}}$ is a critical point of
$\left(\mathbf{I}_{\phi(\partial
\mathsf{K})}^{\varepsilon}(\mathsf{K},\mathds{L}^3);\mathfrak{S}^{\mathsf{K}}\right)$,
where
\begin{itemize}
    \item $\mathbf{I}_{\phi(\partial
\mathsf{K})}^{\varepsilon}(\mathsf{K},\mathds{L}^3)$ is the space of
immersions, $\psi:\mathsf{K}\longrightarrow\mathds{L}^3$, which
satisfy $\psi\vert_{\partial \mathsf{K}}=\phi\vert_{\partial
\mathsf{K}}$; $N_{\psi}\vert_{\partial
\mathsf{K}}=N_{\phi}\vert_{\partial \mathsf{K}}$ and $\langle
N_{\psi},N_{\psi}\rangle=\varepsilon$, and \vspace{2mm}
    \item
    $\mathfrak{S^{\mathsf{K}}}(\psi)=\int_{\mathsf{K}}\,\|dN_{\psi}\|^2\,dA_{\psi}$,
    where $dA_{\psi}$ denotes the element of area of
$(\mathsf{K},\psi^{*}(g))$.
\end{itemize}    \end{definition}

\vspace{1em}

Once we have shown the Gaussian map approach to the $O(2,1)$
Nonlinear Sigma Model, we focus on proving its conformal invariance.

The Willmore functional for free boundary surfaces in the Euclidean
space, \cite{Willmore}, was extended to surfaces with boundary,
\cite{Weiner}. This functional can also be considered for
non-degenerate surfaces in the Lorentz-Minkowski space and so
extended to those with non-null boundary. In particular, we can
define
$\mathfrak{W}:\mathbf{I}_{\Gamma}^{\varepsilon}(M,\mathds{L}^3)\longrightarrow\mathds{R}$
as
\begin{equation}\label{Willmore}
     \mathfrak{W}(\phi)=\int_{M}\,H_{\phi}^2\,dA_{\phi}+\int_{\partial M}\,\kappa_{\phi}\,ds,
\end{equation}
where  $\kappa_{\phi}$ is the geodesic curvature of $\partial M$ in
$(M, \phi^{*}(g))$. This action defines a variational problem
which is invariant under conformal transformations in
$\mathds{L}^3$. The corresponding critical points, which can be
defined similarly to those of the action $\mathfrak{S}$, are
called \textit{Willmore surfaces} (or Willmore surfaces with
prescribed Gauss map along the boundary). The following result
provides a strong relationship between the variational problems
associated with both functionals

\begin{thm} \label{equivalencia}
The two-dimensional $O(2,1)$ Nonlinear Sigma Model,
$\left(\mathbf{I}_{\Gamma}^{\varepsilon}(M,\mathds{L}^3);\mathfrak{S}\right)$,
turns out to be equivalent to the Willmore variational problem
$\left(\mathbf{I}_{\Gamma}^{\varepsilon}(M,\mathds{L}^3);\mathfrak{W}\right)$.
In particular,
\begin{enumerate}
    \item Both have the same critical points. That is,
$\phi\in\mathbf{I}_{\Gamma}^{\varepsilon}(M,\mathds{L}^3)$ is a
solution of the two-dimensional $O(2,1)$ Nonlinear Sigma Model if
and only if $(M,\phi)$ is a Willmore surface.
    \item The two-dimensional $O(2,1)$ Nonlinear Sigma Model is invariant under
conformal changes in the metric of $\mathds{L}^3$.
\end{enumerate}
\end{thm}
\begin{pf}
If $\varepsilon=-1$, then the immersions
$\phi\in\mathbf{I}_{\Gamma}^{-1}(M,\mathds{L}^3)$, provide
Riemannian surfaces, $\phi(M)$, in $\mathds{L}^3$. Consequently,
we can follow, up to slight changes, the proof made in
\cite{Barros1} for surfaces in the Euclidean space.

If $\varepsilon=1$, choose an immersion
$\phi\in\mathbf{I}_{\Gamma}^{1}(M,\mathds{L}^3)$. Then, it
provides a Lorentzian surface, $\phi(M)$, in $\mathds{L}^3$. Given
$\mathsf{K}$ any non-null polygon, the main aim is to
relate the actions
$\mathfrak{S}^{\mathsf{K}},\mathfrak{M}^{\mathsf{K}}:\mathbf{I}_{\phi(\partial
\mathsf{K})}^{-1}(\mathsf{K},\mathds{L}^3)\longrightarrow\mathds{R}$.
Therefore, the first step is to use the formula
(\ref{Gauss-Equation}) and next to get control on the total
Gaussian curvature. To do so, we need a Gauss-Bonnet formula
working on non-null polygons, no matter the causal character of
the boundary pieces. This is made, with details, in the Appendix.
So, after applying this formula, we get
$$\mathfrak{S}^{\mathsf{K}}(\psi)=
4\,\mathfrak{M}^{\mathsf{K}}(\psi)
-6\int_{\partial\mathsf{K}}\,\kappa_{\psi} \,ds-2\sum_{j=1}^r\,\theta_j.$$
Finally, the nature of the hyperbolic angle joint the formula
(\ref{Euler}), (see Appendix), are used to show that the action
measuring the total geodesic curvature of $\psi(\partial
\mathsf{K})=\phi(\partial \mathsf{K})$ indeed does not depend on
$\psi\in \mathbf{I}_{\phi(\partial
\mathsf{K})}^{-1}(\mathsf{K},\mathds{L}^3)$. This concludes the
proof of the result.\end{pf}

\section{Solutions of the $O(2,1)$ NSM which are $\mathbf{A}_{1}$-invariant} \label{Auno-solutiones}

In this section, we completely determine the moduli space of
solutions of the two-dimensional $O(2,1)$ Nonlinear Sigma Model,
$\left(\mathbf{I}_{\Gamma}^{\varepsilon}(M,\mathds{L}^3);\mathfrak{S}\right)$,
which, in addition, are invariant under $\mathbf{A}_{1}$, i.~e., the
group of rotations with time-like axis, $\eje$.

Firstly, we need to establish this previous problem in a suitable
way. In fact, the boundary conditions, $(\Gamma, N_o)$, cannot be
arbitrary but invariant under the $\mathbf{A}_{1}$-action. This
invariance holds if and only if the following conditions are
satisfied:
\begin{enumerate}
    \item The boundary, $\Gamma$, consists of a pair of circles,
    $\{\gamma_1,\gamma_2\}$, contained in Euclidean planes,
    $P_1,P_2$, which are orthogonal to the time-like axis,
    $\eje$, and centered at the points
    $P_i\bigcap\eje$, $i=1,2$.
    \item The unitary normal vector field, $N_o$, along
    $\Gamma=\{\gamma_1,\gamma_2\}$, satisfies $\langle
    N_o,\vec{x}\rangle=\textrm{constant}$.
\end{enumerate}
Consequently, the topology of the surface is $M=[\,a_1,a_2]\times\mathds{S}^1$.

Secondly, the action of $\mathbf{A}_{1}$ on $\mathds{L}^3$ can be
naturally extended to
$\mathbf{I}_{\Gamma}^{\varepsilon}(M,\mathds{L}^3)$ as follows
\[\mathbf{A}_{1}\times\mathbf{I}_{\Gamma}^{\varepsilon}(M,\mathds{L}^3)\longrightarrow\mathbf{I}_{\Gamma}^{\varepsilon}(M,\mathds{L}^3),
\quad (\mu_t,\phi)\mapsto\mu_t\circ\phi.\] It is obvious that both
functionals, $\mathfrak{S}$ and $\mathfrak{W}$ are invariant under
this action, i.~e.,
\[\mathfrak{S}(\mu_t\circ\phi)=\mathfrak{S}(\phi), \,\, \mathfrak{W}(\mu_t\circ\phi)=\mathfrak{W}(\phi),\quad \forall t\in\mathds{R} \,\, \textrm{and} \,\,
 \phi\in\mathbf{I}_{\Gamma}^{\varepsilon}(M,\mathds{L}^3).\]
 Define the set of the immersions which are invariant under $\mathbf{A}_1$, also called symmetric
 points, as
 $\Sigma^{\varepsilon}=\{\phi\in\mathbf{I}_{\Gamma}^{\varepsilon}(M,\mathds{L}^3)\,:\,\mu_t\circ\phi=\phi,\,\forall
 t\in\mathds{R}\}$. To identify $\Sigma^{\varepsilon}$, choose an
 orthonormal basis, $\mathcal{B}=\{\vec{x},\vec{y},\vec{z}\}$, in
 $\mathds{L}^3$ and take the Lorentzian half-plane
 $\mathbf{AdS_2}=\{\vec{v}\in\mathds{L}^3\,:\,\langle\vec{v},\vec{y}\rangle>0,
 \, \langle\vec{v},\vec{z}\rangle=0\}$. Let $\mathbf{C}^{\varepsilon}$ be the
 space of curves, $\alpha:[s_1,s_2]\longrightarrow \mathbf{AdS}_2$, satisfying the
 following conditions, up to a reparametrization:
 \begin{itemize}
    \item $\langle \alpha'(s),\alpha'(s)\rangle=-\varepsilon$,
    \item $\alpha(s_i)=\textrm{trace}(\gamma_i)\cap \mathbf{AdS}_2$, $i=1,2$, and
    \item $\alpha'(s_i)=\nu(\alpha(s_i))$, $i=1,2$.
 \end{itemize}
 Since $M=[\,a_1,a_2]\times \mathds{S}^1$, for each $\alpha\in\mathbf{C}^{\varepsilon}$, we can construct the immersion
 $\phi_{\alpha}\in\mathbf{I}_{\Gamma}^{\varepsilon}(M,\mathds{L}^3)$ defined as  $\phi_{\alpha}(s,e^{it})=\mu_t(\alpha(s))$. It is obvious that $\phi_{\alpha}\in\Sigma^{\varepsilon}$.
The converse also holds. Indeed, given
$\phi\in\Sigma^{\varepsilon}$, we can find
$\alpha\in\mathbf{C}^{\varepsilon}$ such that
$\phi=\phi_{\alpha}$. As a consequence, we can identify
$\Sigma^{\varepsilon}$ with $\mathbf{C}^{\varepsilon}$. On the
other hand,  since $\mathbf{A}_1$ is compact, we can apply the
principle of symmetric criticality \cite{Palais}. According to
this principle, the critical points of $\mathfrak{W}$
(equivalently $\mathfrak{S}$) which are symmetric are just the
critical points of $\mathfrak{W}$ (equivalently $\mathfrak{S}$)
when restricted to $\Sigma^{\varepsilon}$.

The following result provides, up to similarities in
$\mathds{L}^3$, all the solutions of the two-dimensional $O(2,1)$
Nonlinear Sigma Model which are invariant under $\mathbf{A}_1$.

\begin{thm} 
An immersion
$\phi_{\alpha}\in\mathbf{I}_{\Gamma}^{\varepsilon}(M,\mathds{L}^3)$
is a solution of the two-dimensional $O(2,1)$ Nonlinear Sigma Model
if and only if the curve $\alpha$ is a free elastica of
$\mathbf{AdS}_2$ when viewed as an anti de Sitter plane.
\end{thm}

\begin{pf} First, we view a piece of the Lorentz-Minkowski three
space as a warped product:
\[\left(\mathds{L}^3\setminus\eje,g\right)=(\mathbf{AdS}_2,g)\times_{h}(\mathds{S}^1,dt^2),\]
where the warping function,
$h:\mathbf{AdS}_2\longrightarrow\mathds{R}^{+}$, is defined as
$h(p)=\langle\vec{p},\vec{y}\rangle$, where $\vec{p}$ denotes the
position vector of the point $p$, and the metric in
$\mathbf{AdS}_2$ is the induced from the usual one in
$\mathds{L}^3$. Next, we make an obvious conformal change to
obtain a semi-Riemannian product:
$$\left(\mathds{L}^3\setminus\eje,\bar{g}=\frac{1}{h^2}\,g\right)=(\mathbf{AdS}_2,\bar{g})\times(\mathds{S}^1,dt^2).$$
An easy computation shows that $(\mathbf{AdS}_2,\bar{g})$ has
constant Gaussian curvature $-1$, which proves that it is an anti de
Sitter plane.  Denote by $\mathfrak{W}$ and $\overline{\mathfrak{W}}$ the Willmore
functionals of $g$ and $\bar{g}$, respectively. We compute their
restriction to $\Sigma^{\varepsilon}$ as follows
\[\mathfrak{W}(\phi_{\alpha})=\overline{\mathfrak{W}}(\phi_{\alpha})=\int_{M}\,\left(\overline{H}_{\alpha}^{2}+
\overline{R}_{\alpha}\right)\,d\overline{A}_{\alpha}+\int_{\partial
M}\,\overline{\kappa}\,ds,\] where $\overline{R}_{\alpha}$ stands
for the sectional curvature of
$\left(\mathds{L}^3\setminus\eje,\bar{g}\right)$ along $d\phi(TM)$.
Notice that in this case $\overline{R}_{\alpha}=0$ because
$d\phi(TM)$ is a mixed section in a semi-Riemannian product (see \cite{ON}).
Furthermore, the geodesic curvature of $\partial M$ in
$(M,\phi_\alpha ^* (\bar{g}))$, $\overline{\kappa}$, also vanishes
identically since $\Gamma=\phi(\partial M)$ is made up of two
geodesics in $\left(\phi(M),\bar{g}\right)$. Next, we compute the
mean curvature function of $\phi(M)$ in
$\left(\mathds{L}^3\setminus\eje,\bar{g}\right)$, obtaining
\[\overline{H}_{\alpha}^{2}=\frac{1}{4}\overline{\kappa}_{\alpha}^2,\]
where $\overline{\kappa}_{\alpha}$ denotes the curvature function
of $\alpha$ in the anti de Sitter plane
$(\mathbf{AdS}_2,\bar{g})$. As a consequence, we have
\[\mathfrak{W}(\phi_{\alpha})=\overline{\mathfrak{W}}(\phi_{\alpha})=\frac{1}{4}\,\int_{[s_1,s_2]\times\mathds{S}^1}\,
\overline{\kappa}_{\alpha}^2\,ds\,dt=
\frac{\pi}{2}\int_{[s_1,s_2]}\,\overline{\kappa}_{\alpha}^2\,ds.\]
This concludes the proof. \end{pf}

The above result reduces the search of solutions with a pure
rotational symmetry to  curves in the anti de Sitter plane,
$(\mathbf{AdS}_2,\bar{g})$, which are critical points of the
following variational problem, known as the Bernouilli elastica in
its Lorentzian version. The source space is the space of
\textit{clamped} curves, $\mathbf{C}^{\varepsilon}$, and the
Lagrangian is
$\mathfrak{E}:\mathbf{C}^{\varepsilon}\longrightarrow\mathds{R}$,
defined by
$$ \mathfrak{E}(\alpha)=\int_{\alpha}\,\overline{\kappa}_{\alpha}^2\,ds.
$$
The first variation,
$\delta\mathfrak{E}(\alpha):\mathsf{T}_{\alpha}\mathbf{C}^{\varepsilon}\longrightarrow\mathds{R}$,
associated with this functional can be computed, using a standard
method which involves some integration by parts (see for example
\cite{Langer-Singer1} for details) to be,
\[\delta\mathfrak{E}(\alpha)[W]=\int_{\alpha}\bar{g}(\Omega(\alpha),W)\,ds+\left[\mathrm{B}(\alpha,W)\right]_{s_1}^{s_2},\]
where $\Omega$ and $\mathrm{B}$ denote, respectively, the
Euler-Lagrange and the boundary operators, given by
\begin{eqnarray*}
  \Omega(\alpha) &=& 2\varepsilon_2\overline{\nabla}_T^3 T+3\varepsilon_1\overline{\nabla}_T\left(\bar{\kappa}_{\alpha}^2\,T\right)+2\overline{\nabla}_T T, \\
  \mathrm{B}(\alpha,W) &=&
  2\varepsilon_2\bar{g}(\overline{\nabla}_T W,\overline{\nabla}_T
  T)-\bar{g}\left(W,2\varepsilon_2 \overline{\nabla}_T^2
  T+3\varepsilon_1\bar{\kappa}_{\alpha}^2\,T\right),
\end{eqnarray*}
where $\overline{\nabla}$ is the Levi-Civita connection of
$(\mathbf{AdS}_2,\bar{g})$, $\varepsilon_1$ is the causal
character of $T=\alpha'$ and $\varepsilon_2$ that of the normal.
By using the boundary conditions of curves in
$\mathbf{C}^{\varepsilon}$ (clamped curves) we can see that
$\left[\mathrm{B}(\alpha,W)\right]_{s_1}^{s_2}=0$. Therefore, the
elasticae of $(\mathbf{AdS}_2,\bar{g})$ are those curves in
$\mathbf{C}^{\varepsilon}$ satisfying the Euler-Lagrange equation
$\Omega(\alpha)=0$. This equation can be transformed using the
Frenet equations to obtain the following elastica equation
\begin{equation}\label{elastica-Euler-Lagrange}
    2\bar{\kappa}_{\alpha}''-\bar{\kappa}_{\alpha}^3+2\varepsilon_2\,\bar{\kappa}_{\alpha}=0.
\end{equation}
Certainly $\bar{\kappa}_{\alpha}=0$ is a trivial solution of this
equation, which means that geodesics of $(\mathbf{AdS}_2,\bar{g})$
are elasticae. Writting $u=\bar{\kappa}_{\alpha}$, the above
elastica equation turns out to be $2u''-u^3+2\varepsilon_2\,u=0$,
which can be integrated by means of Jacobi elliptic functions,
\cite{Davis}, to have
$$
    u(s)=C\,\mathbf{cn}\left(\lambda(s-a_o),\tilde{C}\right)
$$
where $\lambda\in\mathds{C}\setminus\{0\}$ and $a_o\in\mathds{R}$
are arbitrary constants,
$C^2=-2\left(\lambda^2-\varepsilon_2\right)$ and
$\tilde{C}^2=\frac{\lambda^2-\varepsilon_2}{2\lambda^2}.$ However,
the elliptic cosinus of Jacobi is a complex-valued function, and
the curvature must be a real-valued function. The real-valued
solutions of the equation can be obtained using the properties of
the Jacobi elliptic cosinus, see \cite{Byrd-Friedman}. They
provide the following curvature functions
$$
    \bar{\kappa}_{\alpha}(s)=C\,\mathbf{cn}\left(\sqrt{\varepsilon_2
    -\frac{C^2}{2}}\,(s-a_o),\tilde{C}\right),
$$

$$\textrm{for}\,\,\,  s\in\,\left\{\begin{array}{lr}\mathds{R}\setminus
    \displaystyle\bigcup_{n\in\mathds{Z}}\left\{a_o
    +\frac{2n+1}{\sqrt{\frac{C^2}{2}-\varepsilon_2}}\textsf{E}^{\prime}\right\}
    & \mathrm{ if }\,\,\,\,\,\, \varepsilon_2
    -\frac{C^2}{2}<0 \\
    \\
    \mathds{R} & \mathrm{ if }\,\,\,\,\,\, \varepsilon_2
    -\frac{C^2}{2}>0
    \end{array}\right\},
$$
where $C \in
\,\mathds{R}\setminus\{-\sqrt{2\,\varepsilon_2},\sqrt{2\,\varepsilon_2}\}$
and $a_o\in\mathds{R}$ are arbitrary constants,
$\tilde{C}^2=\frac{C^2}{2 C^2-4\varepsilon_2}$ and
$\textsf{E}^{\prime}$ is the complete elliptic integral of first
kind with modulus $\sqrt{1-\tilde{C}^2}$.

\begin{center}
\begin{tabular}{|c|c|} \hline
&\\
 \includegraphics[scale=1]{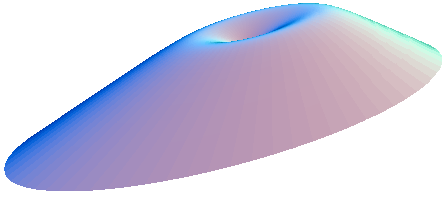} & \includegraphics[scale=1]{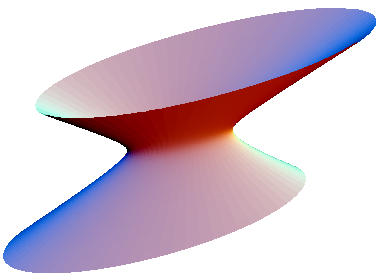}\\
&\\
\hline $\mathbf{A}_{1}$-invariant Riemannian soliton & $\mathbf{A}_{1}$-invariant Lorentzian soliton \\
\hline
\end{tabular}

\end{center}
\section{$\mathbf{A}_{2}$-invariant surfaces in $\mathds{L}^3$} \label{superficiesAdos}

Next, our aim is to obtain, up to isometries of the
Lorentz-Minkowski space, the whole class of solutions of the
$O^1(3)$ Nonlinear Sigma Model, which are invariant under the group
$\mathbf{A}_{2}$. A priori, it could be similar to the above
studied 
case, but it becomes more difficult and subtle. 
The main difficulty we have now, is to find the symmetric
points, i.e. the immersions that are invariant under
$\mathbf{A}_{2}$. 

Let denote by $\eje$ the space-like axis and
consider the only two degenerate planes containing $\eje$.
$\mathds{L}^3$ minus these two planes consists of four open
regions that we will call \textit{fundamental regions}. The
$\mathbf{A}_{2}$-invariant surfaces contained in a fundamental
region will be named \textit{fundamental symmetric surfaces}.

Certainly, we can get a wide class of surfaces invariant under
$\mathbf{A}_{2}$ by taking a curve immersed in any non-degenerate
plane of $\mathds{L}^3$ containing $\eje$, whose trace does not intersect the axis, and rotating it by
applying the elements of $\mathbf{A}_{2}$. These surfaces are
those known in the literature as \textit{rotational surfaces with
space-like axis}, see for example \cite{Hano-Nomizu}. All of them
are fundamental symmetric surfaces. However, we can also find symmetric surfaces that leave a fundamental region to
emerge in another one. In some sense, they are obtained by
gluing fundamental symmetric surfaces. These extended surfaces
have been usually avoided in the literature because of their
difficulty. Nevertheless, the class includes famous surfaces, such
as a saddle surface and a one-sheet hyperboloid. 

In this big
section, we will make an exhaustive analysis to completely
describe the whole class of surfaces in $\mathds{L}^3$ that admit
a rotational group of symmetries with space-like axis, i.e.
surfaces that are invariant under a group of hyperbolic rotations.
For the sake of clearness, we will split our study in several
subsections.

\subsection{Fundamental regions and fundamental surfaces}

Let $\vec{x}$ be a unitary space-like vector in $\mathds{L}^3$. We choose  an orthonormal basis,
$\mathcal{B}=\{\vec{x},\vec{y},\vec{z}\}$, where $\vec{y}$ is also space-like and $\vec{z}$ is
time-like. We work in coordinates with respect to $\mathcal{B}$, so
that the metric in $\Ltres$ is written as $g\equiv dx^2+dy^2-dz^2$.  In $\mathds{L}^3\setminus\eje$ we
will distinguish the following regions that will be called
\textit{fundamental regions}
\begin{eqnarray*}
  \mathscr{R}^{+} &=& \{(x,y,z)\in\mathds{L}^3\, : \, z^2-y^2>0,\, z>0\}, \\
  \mathscr{R}^{-} &=& \{(x,y,z)\in\mathds{L}^3\, : \, z^2-y^2>0,\, z<0\}, \\
  \mathscr{Q}^{+} &=& \{(x,y,z)\in\mathds{L}^3\, : \, z^2-y^2<0,\, y>0\},\,\,\, \mathrm{ and } \\
  \mathscr{Q}^{-} &=& \{(x,y,z)\in\mathds{L}^3\, : \, z^2-y^2<0,\,
  y<0\}.
\end{eqnarray*}

\begin{definition}
An $\mathbf{A}_{2}$-invariant surface immersed in $\Ltres$ is said to be a \emph{fundamental symmetric surface} (or simply a fundamental surface) if it is contained in only one fundamental region.
\end{definition}

We define $\mathscr{R}=\{(x,y,z)\in\mathds{L}^3\, : \, y=0\}$ and
$\mathscr{Q}=\{(x,y,z)\in\mathds{L}^3\, : \, z=0\}$. In this setting, we can introduce the notion of rotational surface generated by a curve. 

\begin{definition} \label{sigmalfa} Let  $\gamma$ be a curve immersed in either
$\mathscr{R}$ or $\mathscr{Q}$,  with domain
$\mathrm{I}\subseteq\mathds{R}$. We define the \emph{rotational surface generated by $\gamma$} as 
$$\mathbf{\Sigma_{\gamma}}:=\{\xi_t(\gamma(s))\, : \, s\in \mathrm{I}, \,
    t\in\mathds{R}\}.$$
\end{definition}

\begin{remark}\label{nota-rotacionales-generadas}
Notice that, in the case $\gamma$ intersects $\eje$, $\mathbf{\Sigma_{\gamma}}$ is not a topological surface.
\end{remark}

Next, we consider the following open half planes
\begin{eqnarray*}
  \tilde{\mathscr{R}}^{+} &=& \mathscr{R}^{+}\cap\mathscr{R}=\{(x,0,z)\in\mathds{L}^3\, : \, z>0\},\\
  \tilde{\mathscr{R}}^{-} &=& \mathscr{R}^{-}\cap\mathscr{R}=\{(x,0,z)\in\mathds{L}^3\, : \, z<0\}, \\
  \tilde{\mathscr{Q}}^{+} &=& \mathscr{Q}^{+}\cap\mathscr{Q}=\{(x,y,0)\in\mathds{L}^3\, : \, y>0\}, \\
  \tilde{\mathscr{Q}}^{-} &=& \mathscr{Q}^{-}\cap\mathscr{Q}=\{(x,y,0)\in\mathds{L}^3\, : \, y<0\},\\
  \mathscr{P}_{+}^{+} &=& \left(\partial\mathscr{R}^{+}\cap\partial\mathscr{Q}^{+}\right)\setminus\eje=\{(x,y,z)\in\mathds{L}^3\, : \,
  y=z>0\},
\\
  \mathscr{P}_{+}^{-} &=& \left(\partial\mathscr{R}^{+}\cap\partial\mathscr{Q}^{-}\right)\setminus\eje=\{(x,y,z)\in\mathds{L}^3\, : \,
  -y=z>0\},
\\
  \mathscr{P}_{-}^{+} &=& \left(\partial\mathscr{R}^{-}\cap\partial\mathscr{Q}^{+}\right)\setminus\eje=\{(x,y,z)\in\mathds{L}^3\, : \, y=-z>0\}, \,\,\,\mathrm{ and }\\
  \mathscr{P}_{-}^{-} &=& \left(\partial\mathscr{R}^{-}\cap\partial\mathscr{Q}^{-}\right)\setminus\eje=\{(x,y,z)\in\mathds{L}^3\, : \,
  y=z<0\}.
  \end{eqnarray*}
  We also consider the following Lorentzian unitary  circles
  \begin{eqnarray*}
    \mathscr{H}^{+} &=& \{(0,y,z)\in\mathds{L}^3\, : \, z^2-y^2=1, \, z>0\},\\
    \mathscr{H}^{-} &=& \{(0,y,z)\in\mathds{L}^3\, : \, z^2-y^2=1,\, z<0\}, \\
    \mathscr{J}^{+} &=& \{(0,y,z)\in\mathds{L}^3\, : \, y^2-z^2=1, \, y>0\}, \,\,\,\mathrm{ and } \\
    \mathscr{J}^{-} &=& \{(0,y,z)\in\mathds{L}^3\, : \, y^2-z^2=1, \,
    y<0\}.
  \end{eqnarray*}
  It should be noticed that while $\mathscr{H}^{+}$ and
  $\mathscr{H}^{-}$ are space-like in $\mathds{L}^3$, with metric $dt^2$,
  $\mathscr{J}^{+}$ and $\mathscr{J}^{-}$ are time-like, with metric
  denoted by $-dt^2$. Next, we define the following positive functions
  \begin{eqnarray*}
    f_{+}:\tilde{\mathscr{R}}^{+}\longrightarrow\mathds{R}, \quad f_{+}(x,0,z) &=& z ,\\
    f_{-}:\tilde{\mathscr{R}}^{-}\longrightarrow\mathds{R}, \quad f_{-}(x,0,z) &=& -z ,\\
    h_{+}:\tilde{\mathscr{Q}}^{+}\longrightarrow\mathds{R}, \quad h_{+}(x,y,0) &=& y ,\,\,\,\mathrm{ and }\\
    h_{-}:\tilde{\mathscr{Q}}^{-}\longrightarrow\mathds{R}, \quad h_{-}(x,y,0)
    &=&-y.
  \end{eqnarray*}
  In this setting, it is not difficult to check the following warped product decompositions
  \begin{eqnarray*}
    (\mathscr{R}^{+},g) &=& (\tilde{\mathscr{R}}^{+},g)\times_{f_{+}}(\mathscr{H}^{+},dt^2), \\
    (\mathscr{R}^{-},g) &=& (\tilde{\mathscr{R}}^{-},g)\times_{f_{-}}(\mathscr{H}^{-},dt^2), \\
    (\mathscr{Q}^{+},g) &=& (\tilde{\mathscr{Q}}^{+},g)\times_{h_{+}}(\mathscr{J}^{+},-dt^2), \,\,\,\mathrm{ and } \\
    (\mathscr{Q}^{-},g) &=&
    (\tilde{\mathscr{Q}}^{-},g)\times_{h_{-}}(\mathscr{J}^{-},-dt^2).
  \end{eqnarray*}
  Furthermore, when we make the obvious conformal changes, it is
  easy to see that:
  \begin{itemize}
    \item $\left(\tilde{\mathscr{R}}^{+},\frac{1}{f_{+}^2}g\right)$
    and $\left(\tilde{\mathscr{R}}^{-},\frac{1}{f_{-}^2}g\right)$
    are de Sitter planes with curvature $1$, and
    \item $\left(\tilde{\mathscr{Q}}^{+},\frac{1}{h_{+}^2}g\right)$
    and $\left(\tilde{\mathscr{Q}}^{-},\frac{1}{h_{-}^2}g\right)$
    are hyperbolic planes with curvature $-1$.
  \end{itemize}
  Consequently, we obtain the following result.
\begin{lem} \label{warped-product}
\begin{enumerate}
    \item $\left(\mathscr{R}^{+},\frac{1}{f_{+}^2}g\right)$ and $\left(\mathscr{R}^{-},\frac{1}{f_{-}^2}g\right)$ are the
    semi-Riemannian product of a de Sitter plane and a space-like Lorentzian unitary
    circle.
    \item $\left(\mathscr{Q}^{+},\frac{1}{h_{+}^2}g\right)$ and $\left(\mathscr{Q}^{-},\frac{1}{h_{-}^2}g\right)$ are the
    semi-Riemannian product of a hyperbolic plane and a time-like Lorentzian unitary
    circle.
  \end{enumerate}
\end{lem}

\subsection{Fundamental symmetric immersions}
In this subsection, we completely describe those non-degenerate immersions,
$\phi:M\longrightarrow\mathds{L}^3$, whose image is a fundamental symmetric surface. 
We will see that they 
correspond with rotational surfaces generated by non-degenerate curves that do not intersect $\eje$.
To proceed with, we consider separately Riemannian and Lorentzian cases. 

Since $\mathbf{A}_2$-invariant Riemannian surfaces
automatically lie in $\mathscr{R}^{+}$ or $\mathscr{R}^{-}$, the following result assures us that all the $\mathbf{A}_2$-invariant Riemannian surfaces are fundamental symmetric surfaces, and it classifies them.

\begin{thm} \label{Riemannianas-Invariantes} Let $M$ be a connected surface and $\phi:M\longrightarrow\mathds{L}^3$ an immersion. Then,
 $(M,\phi^{*}(g))$ is Riemannian and $\mathbf{A}_2$-invariant if and only if
there exists a smooth time-like curve, $\alpha$, contained in either $\tilde{\mathscr{R}}^{+}$ or $\tilde{\mathscr{R}}^{-}$, such that $\phi(M)=\Sigma_{\alpha}$. In particular, these surfaces lie in $\mathscr{R}^{+}$ or $\mathscr{R}^{-}$.
\end{thm}

\begin{pf} The sufficient condition is widely known,
\cite{Hano-Nomizu}. To prove the converse, assume that
$\phi:M\longrightarrow\mathds{L}^3$ is an immersion such that
$(M,\phi^{*}(g))$ is Riemannian and $\phi(M)$ is invariant under
the action of $\mathbf{A}_2$. This implies that $\phi(M)$ is
foliated by space-like orbits. On the other hand, the orbits in
$\mathscr{Q}^{+}\bigcup\mathscr{Q}^{-}$ are time-like and those in
$\mathscr{P}_{+}^{+}\bigcup\mathscr{P}_{+}^{-}\bigcup\mathscr{P}_{-}^{+}\bigcup\mathscr{P}_{-}^{-}$
are light-like. This shows that
$\phi(M)\subset\mathscr{R}^{+}\bigcup\mathscr{R}^{-}\bigcup\eje$.
Consequently, there exists a space-like curve,
$\alpha:J\subset\mathds{R}\longrightarrow\mathscr{R}$,  such that
$\phi(M)=\{\xi_t(\alpha(s))\, : \, s\in J, \, t\in\mathds{R}\}$.
However, if $\alpha(J)\bigcap\tilde{\mathscr{R}}^{+}\neq\emptyset$
then $\alpha(J)\subset\tilde{\mathscr{R}}^{+}$. Indeed, given a
point $s_0\in J$ such that $\alpha(s_0)\in \eje$, then the
surface $\Sigma_{\alpha}$ is not even a topological manifold at
$\alpha(s_0)$.  This concludes the proof.
\end{pf}

\vspace{1em}

In the Lorentzian case, the behavior is different. The reason is that it is possible to find $\mathbf{A}_2$-invariant Lorentzian surfaces in the four fundamental regions, according to the following result.

\begin{thm} \label{Lorentzianas-Invariantes-Fundamentales}
Let $M$ be a connected surface and $\immersion$. 
Then, $(M,\phi^{*}(g))$ is a Lorentzian symmetric fundamental surface if and only if either
\begin{itemize}
    \item  
    there exists a time-like
    curve, $\alpha:\mathrm{I}\longrightarrow\tilde{\mathscr{R}}^{+}$, such that $\phi(M)=\Sigma_{\alpha}$,
    \item  
    there exists a time-like
    curve, $\alpha:\mathrm{I}\longrightarrow\tilde{\mathscr{R}}^{-}$, such that $\phi(M)=\Sigma_{\alpha}$,
    \item  
    there exists a space-like
    curve, $\alpha:\mathrm{I}\longrightarrow\tilde{\mathscr{Q}}^{+}$, such that $\phi(M)=\Sigma_{\alpha}$,  or 
    \item  
    there exists a space-like
    curve, $\alpha:\mathrm{I}\longrightarrow\tilde{\mathscr{Q}}^{-}$, such that $\phi(M)=\Sigma_{\alpha}$.
\end{itemize}
\end{thm}
The proof is left to the reader because it is similar to the one of Theorem \ref{Riemannianas-Invariantes}.

\subsection{Some examples to motivate the extended Lorentzian case}\label{examples-motivation}

For a better understanding of the  general (or extended)
Lorentzian case, we analyze two examples of
$\mathbf{A}_2$-invariant Lorentzian surfaces that intersect more than one fundamental region.

\vspace{1em}

\noindent \textbf{Example 1 : A saddle surface.} We consider the following saddle surface in $\Ltres$
\[\mathbf{S}=\left\{(x,y,z)\in\mathds{L}^3 \, : \,
x=y^2-z^2>-\frac{1}{4}\right\},\]
which admits a natural Monge parametrization as a graph. Indeed, in the plane $x=0$, we  consider the map
$$\mathrm{X}:\mathds{R}\times\left(-\frac{1}{2},\frac{1}{2}\right)\longrightarrow \mathbf{S}\subset\mathds{L}^3,
\quad \mathrm{X}(y,z)=(y^2-z^2,y,z).$$ $\mathbf{S}$ is a Lorentzian
surface in $\mathds{L}^3$, because we have considered only the piece
where the induced metric is Lorentzian. $\mathbf{S}$ is also invariant under
the group $\mathbf{A}_2$. In addition, every fundamental region
contains a piece of this saddle surface. According to the notation we
are using, these pieces can be described as follows:
\begin{eqnarray*}
  \Sigma_{\alpha^{+}}, & \mbox{ where } &\alpha^{+}:\left(0,1/2\right)\rightarrow\tilde{\mathscr{R}}^{+}, \,\,\alpha^{+}(s)=(-s^2,0,s), \\
  \Sigma_{\alpha^{-}}, & \mbox{ where } &\alpha^{-}:\left(-1/2,0\right)\rightarrow\tilde{\mathscr{R}}^{-}, \,\,\alpha^{-}(s)=(-s^2,0,s), \\
  \Sigma_{\beta^{+}}, & \mbox{ where } &\beta^{+}:\left(0,+\infty\right)\rightarrow\tilde{\mathscr{Q}}^{+}, \,\,\beta^{+}(s)=(s^2,s,0), \,\,\mathrm{ and }\\
 \Sigma_{\beta^{-}}, & \mbox{ where } &\beta^{-}:\left(-\infty,0\right)\rightarrow\tilde{\mathscr{Q}}^{-}, \,\,\beta^{-}(s)=(s^2,s,0).
\end{eqnarray*}
Obviously, these surfaces are glued along the common boundaries,
obtaining $\mathbf{S}$. Though the gluing mechanism is obvious in
this case, we will emphasize it as a motivation for the later
extension. We will work in a neighborhood of the boundaries of the
above four pieces. 

Firstly, it should be noticed that we can work with the following couple
of curves:
\begin{itemize}
    \item a time-like curve
    $\alpha:(-\delta,\delta)\longrightarrow\mathscr{R}$,
    $\alpha(s)=(f_{\alpha}(s),0,s)=(-s^2,0,s)$, and
    \item a space-like curve
    $\beta:(-\delta,\delta)\longrightarrow\mathscr{Q}$,
    $\beta(s)=(f_{\beta}(s),s,0)=(s^2,s,0)$.
\end{itemize}
They are defined as graphs for a certain $\delta\in(0,1/2)$. In
addition, we have a gluing smooth function, $F:\{(y,z)\in\mathds{R}^2\, :
\, |z^2-y^2|<\delta^2\}\longrightarrow\mathds{R}$, defined as
\[F(y,z)=\left\{\begin{array}{cll} f_{\alpha}\left(\textrm{sign}(z)\sqrt{z^2-y^2}\right)=y^2-z^2& \quad \textrm{if} & \quad z^2\geq y^2,\\
f_{\beta}\left(\textrm{sign}(y)\sqrt{y^2-z^2}\right)=y^2-z^2&
\quad \textrm{if}& \quad y^2\geq z^2.\end{array} \right\} \]
When we consider the four pieces altogether, the Monge parametrization is just obtained in
terms of $F$.

\vspace{2mm}

\noindent \textbf{Example 2 : A one-sheet hyperboloid.} We consider the following one-sheet hyperboloid
$$
    \mathbf{H}=\{(x,y,z)\in\mathds{L}^3 \, : \,
    x^2+y^2-z^2=1\}.
$$
Clearly, it is $\mathbf{A}_2$-invariant and it is not contained in any
fundamental region.  In fact, the intersection of $\mathbf{H}$ and the fundamental regions consists of six connected pieces. We denote by $p = (1, 0, 0)$ and $q=(-1, 0, 0)$ the two points in which $\mathbf{H}$ intersects the axis. Then,  it is easy to check that the boundaries of above six pieces are just the eight light-like orbits with boundary either $p$ or $q$. Thus, it is necessary to glue twice to obtain $\mathbf{H}$.

Firstly, we work around $p$. We choose $\delta$ satisfying $0<\delta<1$ and we define:
\begin{itemize}
    \item a time-like curve $\alpha_p:(-\delta,\delta)\rightarrow\mathscr{R}$,
    $\alpha_p(s)=(f_{\alpha_p}(s),0,s)=(+\sqrt{1+s^2},0,s)$, and
    \item a space-like curve $\beta_p:(-\delta,\delta)\rightarrow\mathscr{Q}$,
    $\beta_p(s)=(f_{\beta_p}(s),s,0)=(+\sqrt{1-s^2},s,0)$,
\end{itemize}
which satisfy $\alpha_p(0)=\beta_p(0)=p$. Then, we define the gluing
smooth function $\mathbf{F_p}:\{(y,z)\in\mathds{R}^2\, : \,
|z^2-y^2|<\delta^2\}\longrightarrow\mathds{R}$ as
\[\mathbf{F_p}(y,z)=\left\{\begin{array}{lll} f_{\alpha_p}\left(\textrm{sign}(z)\sqrt{z^2-y^2}\right)=+\sqrt{1-y^2+z^2}& \quad \textrm{if} & \quad z^2\geq y^2,\\
f_{\beta_p}\left(\textrm{sign}(y)\sqrt{y^2-z^2}\right)=+\sqrt{1-y^2+z^2}&
\quad \textrm{if}& \quad y^2\geq z^2.\end{array} \right\} \] Now, in
terms of this gluing function, we can define a parametrization of
the one-sheet hyperboloid around $p$ as follows
$$\mathrm{X_p}:\{(y,z)\in\mathds{R}^2\, : \,
|z^2-y^2|<\delta^2\}\longrightarrow\mathbf{H}\subset\mathds{L}^3,
 $$
 $$\mathrm{X_p}(y,z)=\left(\mathbf{F_p}(y,z),y,z\right)=\left(+\sqrt{1-y^2+z^2},y,z\right).$$
Finally, using the negative square root we obtain curves and a
gluing function to paste the pieces around $q$.

\begin{center}
\begin{tabular}{|c|c|} \hline & \\
\includegraphics[scale=.7]{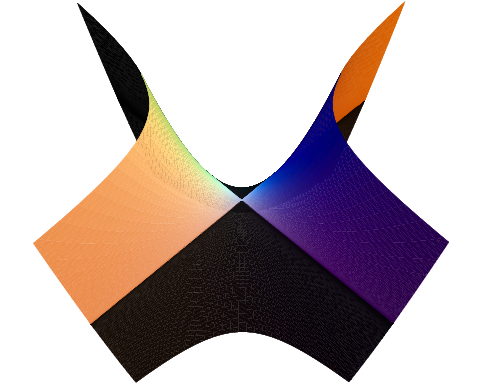}
& \includegraphics[scale=.5]{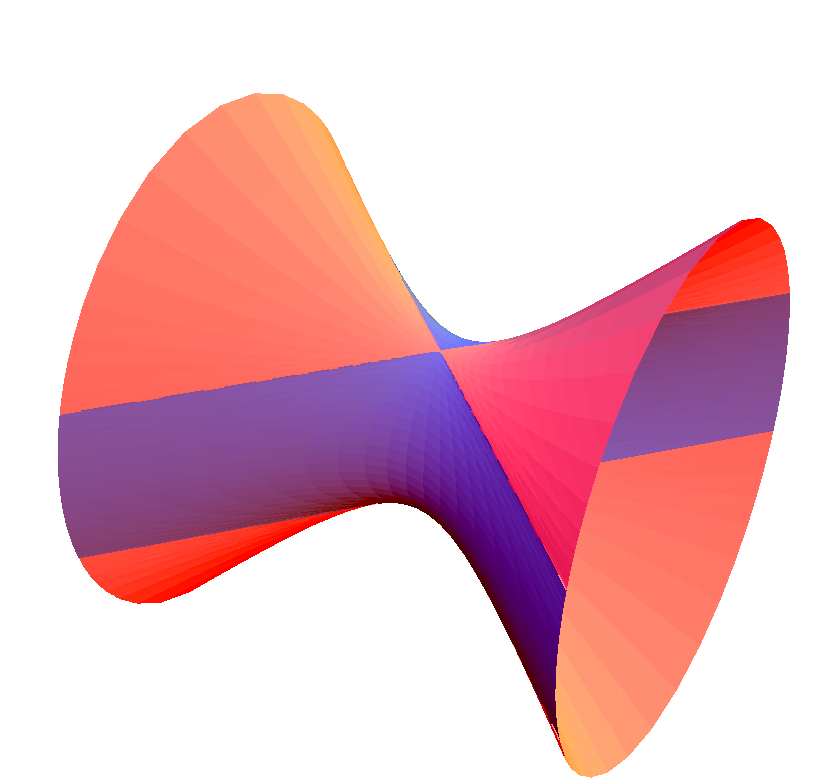}
\\
&\\
\hline
Saddle surface & One-sheet hyperboloid  \\
\hline
\end{tabular}

\end{center}



\subsection{Dissection of an $\mathbf{A}_2$-invariant Lorentzian surface} \label{diseccion}

Along this subsection, we assume that $M$ is a connected smooth
surface and $\immersion$ is an immersion such that
$(M,\phi^{*}(g))$ is Lorentzian and $\phi(M)$ is
$\mathbf{A}_2$-invariant. We are going to use surgery to study the
pieces of $\phi(M)$ that lie in each of the fundamental regions
and in
$\mathscr{P}_{+}^{+}\bigcup\mathscr{P}_{+}^{-}\bigcup\mathscr{P}_{-}^{+}\bigcup\mathscr{P}_{-}^{-}$.
The following assertions can be checked by the reader.

\begin{enumerate}
    \item
    $\phi(M)\bigcap\left(\mathscr{R}^{+}\bigcup\mathscr{R}^{-}\right)$
    is empty or it is a countable union of Lor\-entzian fundamental
    surfaces that are generated by time-like curves immersed in either $\tilde{\mathscr{R}}^+$ or $\tilde{\mathscr{R}}^-$, i.e.,
    $$\left(\bigcup_{e\in\mathcal{E}}\Sigma_{\alpha_{e}}\right)\bigcup
    \left(\bigcup_{f\in\,\mathcal{F}}\Sigma_{\alpha_{f}}\right),$$
    where $\{\alpha_{e}\, : \, e\in\mathcal{E}\}$  and $\{\alpha_{f}\, : \,
    f\in\mathcal{F}\}$ are
    countable families of time-like curves in
    $\tilde{\mathscr{R}}^+$ and $\tilde{\mathscr{R}}^-$, respectively.
    \item
    $\phi(M)\bigcap\left(\mathscr{Q}^{+}\bigcup\mathscr{Q}^{-}\right)$
    is empty or it is a countable union of Lorentzian fundamental
    surfaces with profile curves immersed in either $\tilde{\mathscr{Q}}^+$ or $\tilde{\mathscr{Q}}^-$, i.e.,
    \[\left(\bigcup_{\lambda\in\,\Lambda}\Sigma_{\beta_{\lambda}}\right)\bigcup
    \left(\bigcup_{\theta\in\Theta}\Sigma_{\beta_{\theta}}\right), \]
    where $\{\beta_{\lambda}\, : \, \lambda\in\Lambda\}$ and
    $\{\beta_{\theta}\, : \, \theta\in\Theta\}$ are countable families of curves in
    $\tilde{\mathscr{Q}}^+$ and $\tilde{\mathscr{Q}}^-$, respectively.
    \item
    $\phi(M)\bigcap\left(\mathscr{P}_{+}^{+}\bigcup\mathscr{P}_{+}^{-}\bigcup\mathscr{P}_{-}^{+}\bigcup\mathscr{P}_{-}^{-}\right)$
     is empty or  a
    countable set of light-like orbits lying in the boundary of the
    Lorentzian fundamental surfaces mentioned in the previous items.
\end{enumerate}

We have already studied the Lorentzian fundamental surfaces, so we will focus on the case in which $\phi(M)$ is not a fundamental surface. So, we assume that $\phi(M )\cap (\mathscr{R}^+\cup\mathscr{R}^-)\neq\emptyset$ and $\phi(M)\cap (\mathscr{Q}^+\cup\mathscr{Q}^-)\neq \emptyset$.

Once we have made the dissection, we will study the curves immersed in $\mathscr{R}$, the curves immersed in $\mathscr{Q}$ and finally, how these curves are related.

\vspace{1em}

\subsubsection*{Generating curves immersed in $\mathscr{R}$}\label{Simetricos-curvas-en-R}

By using a
connection argument, as well as the non existence of closed
time-like curves in $\mathscr{R}$, we can state the following
facts about the curves in $\tilde{\mathscr{R}}^+$ and
$\tilde{\mathscr{R}}^-$:

\begin{enumerate}

\item[{[R1]}] For each $\alpha\in\{\alpha_{e}\, : \, e\in\mathcal{E}\}\cup\{\alpha_{f}\, : \,
    f\in\mathcal{F}\}$, there exists $U_{\alpha}\subseteq M$ connected submanifold such that $\phi(U_{\alpha})=\Sigma_{\alpha}$. Moreover, if there exist more than one of such submanifolds, we include in $\{\alpha_{e}\, : \, e\in\mathcal{E}\}\cup\{\alpha_{f}\, : \,
    f\in\mathcal{F}\}$ as many copies of $\alpha$ as existing submanifolds, and we notate them with different subindices.

\item[{[R2]}] If $\alpha\in\{\alpha_{e}\, : \, e\in\mathcal{E}\}\cup\{\alpha_{f}\, : \,
    f\in\mathcal{F}\}$, then $\alpha$ is maximal in the sense that there are no connected submanifolds satisfying  $U_{\alpha}\subseteq V$ and $\phi(V)$ do not intersect $\eje$.

\item[{[R3]}]  Many curves of
$\{\alpha_e:e\in\mathcal{E}\}\cup\{\alpha_f:f\in\mathcal{F}\}$ can
be glued to obtain smooth or piecewise smooth time-like curves in
$\mathscr{R}$. Indeed, given $p\in \eje$, if there 
exist $e\in\mathcal{E}$ and $f\in\mathcal{F}$ satisfying
\begin{itemize}
 \item $p$
belongs to the boundary of these two curves, and
\item there exists $U\subseteq M$ open and connected, such that $U\cap\phi^{-1}(\mathscr{R}^+)=U_{\alpha_e}$ and $U\cap\phi^{-1}(\mathscr{R}^-)=U_{\alpha_f}$,
\end{itemize}
then these two curves can be glued. If $p\in\phi(M)$, then the
union is a time-like smooth curve in
$\mathscr{R}$. Otherwise, the union is time-like and smooth everywhere
except in $p$, where we only know it is continuous. 

It is easy to check that this procedure cannot be applied to two curves of $\{\alpha_e:e\in\mathcal{E}\}$ or two
curves of $\{\alpha_f:f\in\mathcal{F}\}$, because we will not obtain a Lorentzian surface. 

\item[{[R4]}] After  all the possible gluing processes, we obtain a countable family of continuous piecewise smooth time-like
curves in $\mathscr{R}$, $\{\alpha_i:\mathrm{J}_i\longrightarrow\mathds{R}\,:i\in\mathcal{I}\}$, $\mathrm{J}_i\subseteq \mathds{R}$ being an interval for all $i\in\mathcal{I}$ (they are
smooth everywhere, except in those points  satisfying
$p\notin\phi(M)$ and $p\in\eje$).

\item[{[R5]}] As $\mathscr{R}$ is a Lorentzian plane, we can deduce that for each $i\in\mathcal{I}$ there exists a unique $s_i\in\textrm{Closure}(\mathrm{J}_i)$ such that $p_i=\lim_{s\rightarrow s_i}\alpha(s)$ belongs to $\eje$. In addition, either $s_i$ belongs to the boundary of $\mathrm{J}_i$, or the curve changes from one fundamental region to another at $p_i$.

\end{enumerate}

\vspace{1em}

\subsubsection*{Generating curves immersed in $\mathscr{Q}$}\label{Simetricos-curvas-en-Q}
The following assertions hold:

\begin{enumerate}
\item[{[Q1]}]
The properties analogous to [R1], [R2] and [R3] hold true for curves in  $\{\beta_{\lambda}:\lambda\in\Lambda\}\cup\{\beta_{\theta}:\theta\in\Theta\}$.

\item[{[Q2]}] After  all the possible gluing processes, we obtain only one continuous, piecewise
smooth curve in $\mathscr{Q}$ (smooth everywhere except in the intersections with $\eje$, in which we only know the curve is continuous). We will notate this curve as $\beta$. This property is a consequence of the connectedness of $M$ and [R5]. 

\item[{[Q3]}] The domain of $\beta$, $\mathrm{J}$, can be an interval or $\mathds{S}^1$. The firs situation corresponds to the case in which either $\beta$ is not closed, or $\beta$ is closed but there are two curves in $\{\beta_{\lambda}:\lambda\in\Lambda\}\cup\{\beta_{\theta}:\theta\in\Theta\}$ that suffered only one gluing process.

\item[{[Q4]}] Each time $\beta$ intersects $\eje$, either the curve changes from a fundamental region to another, or the intersection is a boundary point of $\beta$, or the point belongs to the only two curves in $\{\beta_{\lambda}:\lambda\in\Lambda\}\cup\{\beta_{\theta}:\theta\in\Theta\}$ that suffered only one gluing process.

\end{enumerate}

\subsubsection*{Connecting profile curves with different causal character}
At this point, we know that the surface $\phi(M)$ is generated
by $\{\alpha_{i}\, : \, i\in\mathcal{I}\}$ and $\beta$. However,
we need a deeper understanding of the relation between $\beta$ and the
curves $\{\alpha_i:i\in\mathcal{I}\}$, as well as the way of
constructing the original surface from the generating curves.

Thanks to the connectedness of $M$, the Remark \ref{nota-rotacionales-generadas} and the properties above, is easy to check the following assertions.

\begin{description}
\item[\textbf{P1.}] $\forall s_o \in \mathrm{J}$ such that $\beta(s_o)\in\eje$, there exist $\varepsilon>0$ and $i\in\mathcal{I}$, satisfying  
\begin{itemize} 
\item $\beta(s_o)\in\mathrm{Closure}(\traza{\alpha_i})$, and
\item $\Sigma_{\beta_{|\mathrm{J}_o}}\bigcup\Sigma_{\alpha_i}$ is a smooth surface, where $\mathrm{J}_o=]s_o-\varepsilon,s_o+\varepsilon[\,\cap\,\mathrm{J}$.
\end{itemize}
Roughly speaking, for each $s_o \in \mathrm{J}$ such that $\beta(s_o)\in\eje$, there exists $\alpha_i$ gluing appropriately with $\beta$ in a neighborhood of $s_o$.
\item[\textbf{P2.}] $\forall i \in \mathcal{I}$, there exist $\varepsilon>0$ and $s_o\in\mathrm{Closure}(\mathrm{J})$, such that  
\begin{itemize}
\item $\lim_{s\to s_o}\beta(s)=\lim_{s\to s_i}\alpha(s)$, and
\item $\Sigma_{\beta_{|\mathrm{J}_o}}\bigcup\Sigma_{\alpha_i}$ is a smooth surface, where  $\mathrm{J}_o=]s_o-\varepsilon,s_o+\varepsilon[\,\cap\,\mathrm{J}$.
\end{itemize}
Roughly speaking, for each $\alpha_i$, there exists $s_o\in\mathrm{Closure}(\mathrm{J})$ such that $\beta$ and $\alpha_i$ glue appropriately in a neighborhood of $s_o$.
\end{description}

\begin{remark}
If we ask $U_{\alpha_i^+}\cup U_{\alpha_i^-}\cup U_{\beta_i^+}\cup U_{\beta_i^-}$ to be connected, being $\alpha_i^{\pm}=\alpha_i |_{\mathrm{J}_i \cap \mathds{R}^{\pm}}$ and  $\beta^{\pm}=\beta |_{\mathrm{J}_o \cap \mathds{R}^{\pm}}$, then $i$ is unique in $\mathbf{P1}$, and $s_o$ is unique in $\mathbf{P2}$. In this case, $\mathbf{P1}$ and $\mathbf{P2}$ can be viewed as injective maps, $\mathbf{P1}:\{s_o \in\mathrm{J} : \beta(s_o) \in\left\langle  \vec{x}\right\rangle \}\longrightarrow \mathcal{I}$ and $\mathbf{P2}:\mathcal{I}\longrightarrow\{s_o \in\mathrm{Closure}(\mathrm{J}) : \lim_{s\rightarrow s_o}\beta(s) \in\left\langle  \vec{x}\right\rangle \}$, verifying $\mathbf{P2}\circ\mathbf{P1}(s)=s$ for all $s\in \mathrm{J}$ such that $\beta(s)\in\eje$.
\end{remark}

\begin{remark}
When considering $\Sigma_{\beta_{|\mathrm{J}_o}}\cup\Sigma_{{\alpha_i}}$, we are also adding the light-like orbits belonging to the boundary of these two surfaces and to $\phi(M)$ (otherwise the union will not be connected).
\end{remark}

\subsubsection*{Light-like orbits connecting the pieces}

\begin{definition}\label{orbital}
Given any point $p=(x_o,0,0)\in\eje$, we define the {\em light-like orbital} at $p$,
$\mathbf{O}_p$, as
the set consisting of $p$ together with the four light-like orbits through this point, i.e.
$\mathbf{O}_p=\{(x_o,y,z):
    y^2=z^2\}$.
\end{definition}

\begin{definition}\label{ligh-like-patch}For each $s_o\in\mathrm{Closure}(\mathrm{J})$ such that $p=\lim_{s\to s_o}\beta(s)\in\eje$,
we define the {\em light-like patch} at 
$s_o$ as $$B_{s_o}=\mathbf{O}_{p_o}\bigcap\phi(U),$$ where $p_o=\lim_{s\to s_o}\beta(s)$ and $U\subseteq M$ is a domain such that: 
\begin{itemize}
\item $\phi(U)\bigcap \left(\mathscr{Q}^+
\bigcup \mathscr{Q}^-\right)=\Sigma_{\beta_{\mid\mathrm{J}_o}}$, and  
\item if $\exists i\in\mathcal{I}$ satisfying $\mathbf{P2}(i)=s_o$, then $\phi(U)\bigcap\left(\mathscr{R}^+
\bigcup \mathscr{R}^-\right)=\Sigma_{\alpha_i}$.
\end{itemize}
\end{definition} 

\begin{remark}
The light-like patch at $s_o$ is just the union of the elements of a subset of $\{\{p_o\}, \{(x_o,a,a)\,:\,a>0\}, \{(x_o,a,a)\,:\,a<0\}, \{(x_o,-a,a)\,:\,a>0\}, \{(x_o,-a,a)\,:\,a<0\} \}$ where $p_o=(x_o,0,0)=\lim_{s\rightarrow s_o}\beta(s)$.
\end{remark}

\subsubsection*{Summary of the dissection}

Given an $\mathbf{A}_2$-invariant Lorentzian immersion, $\phi:M\longrightarrow \Ltres$, there exists a family of time-like curves in $\mathscr{R}$, $\{\alpha_i:\mathrm{J}_i\subseteq\mathds{R}\longrightarrow\mathscr{R} : i\in\mathcal{I}\}$, and a curve in $\mathscr{Q}$, $\beta:\mathrm{J}\longrightarrow\mathscr{Q}$, satisfying:
\begin{itemize}
 \item All of them are smooth everywhere except in those points belonging to $\eje\backslash \phi(M)$, in which the curve is only known to be continuous.
\item For each $i\in\mathcal{I}$, $\mathrm{J}_i$ is an interval. Even more, there exists only one  $s_i\in\mathrm{Closure}(\mathrm{J}_i)$ such that $p_i=\lim_{s\rightarrow s_i}\alpha(s)$ belongs to $\eje$. If $s_i$ is not a boundary point of $\mathrm{J}_i$, then it is a point in which the curve goes from one fundamental region into another. 
\item The domain of $\beta$, $\mathrm{J}$, is either an interval, or $\mathds{S}^1$, see [Q3].
\item Each time $\beta$ intersects $\eje$, either the curve goes from one fundamental region to another, or the point belongs to $\{\displaystyle\lim_{s\rightarrow s_o}\beta(s)\,/\,s_o\in\partial\mathrm{J}\}$.
\item $\mathbf{P1}$ and $\mathbf{P2}$ hold.
\end{itemize}
Also, for each $s_o\in\mathrm{Closure}(\mathrm{J})$ such that $\lim_{s\rightarrow s_o}\beta(s)\in\eje$, there exists a set, $B_{s_o}$, called light-like patch, consisting of the union of some of the following sets: $\{p_o\}$, $\{(x_o,a,a)\,:\,a>0\}$, $\{(x_o,a,a)\,:\,a<0\}$, $\{(x_o,-a,a)\,:\,a>0\}$ and $\{(x_o,-a,a)\,:\,a<0\}$ (where  $p_o=(x_o,0,0)=\lim_{s\rightarrow s_o}\beta(s)$). 

In this setting,  $$\phi(M)=\Sigma_{\beta}\bigcup\left(\bigcup_{i\in\mathcal{I}}\Sigma_{\alpha_i}\right)\bigcup\left(\bigcup\{B_{s_o}\,:\,\lim_{s\rightarrow s_o}\beta(s)\in\eje\}\right).$$

Note that
$\Sigma_{\alpha_i}$ and $\Sigma_{\beta}$ correspond to Definition \ref{sigmalfa}, but
here we are removing the points of the axis. From now on, we will
use this assumption freely.

\subsection{Characterization of the gluing} \label{orbita-temporal} In this subsection, we
characterize the way to paste two Lorentzian fundamental symmetric surfaces,
$\Sigma_{\alpha}$ and $\Sigma_{\beta}$, which are generated by
suitable curves, $\alpha$ in $\mathscr{R}$ and $\beta$ in
$\mathscr{Q}$. 

Let $\alpha$ be a time-like curve in $\mathscr{R}$ and  $\beta$
 a curve in $\mathscr{Q}$, such that $\lim_{s\rightarrow 0}\alpha(s)=\lim_{s\rightarrow
0}\beta(s)=p=(x_o,0,0)\in\,\eje$. Firstly, we choose an appropriate light-like patch,
$B_0\subset \mathbf{O}_p=\{(x_o,y,z):y^2=z^2\}$. Secondly, we take a neighborhood of $0$ in the domain of $\beta$ such that $\lim_{s\rightarrow \tilde{s}}\beta(s)\notin \eje$, for $\tilde{s}\neq s_o$. There is no loss of generality, because only that neighborhood is important in the gluing process.

In this setting, the following result can be regarded as the master piece to understand how $\mathbf{A}_2$-invariant surfaces generated by curves glue smoothly. Moreover, it characterizes the gluing mechanism.

\begin{thm}\label{Pegado} \textbf{Local Gluing Theorem}
In the setting of this subsection, $\Sigma_\alpha$ and
$\Sigma_\beta$ glue smoothly and the metric along the union is
Lorentzian, i.e. $\Sigma=\Sigma_\alpha\bigcup\Sigma_\beta\bigcup
B_0$ is a smooth Lorentzian surface in a neighborhood of
$B_0$, if, and only if, there exist smooth functions
$f_{\alpha}$ and $f_{\beta}$ such that the following assertions
hold:
\begin{enumerate}[LG1.]
\item \label{LG1} $\alpha(u)=\left(f_{\alpha}(u),0,u\right)$ is a
parametrization of $\alpha$ in a neighborhood of $p$.
\item \label{LG2} $\beta(u)=\left(f_{\beta}(u),u,0\right)$ is a
parametrization of $\beta$ in a neighborhood of $p$.
 \item \label{LG3} The following function is smooth
\[F(y,z)=\left\{\begin{array}{lll}
f_{\alpha}\left(\textrm{sign}(z)\sqrt{z^2-y^2}\right)& \quad
\textrm{if} & \quad z^2\geq y^2,\\
f_{\beta}\left(\textrm{sign}(y)\sqrt{y^2-z^2}\right)& \quad
\textrm{if}& \quad y^2\geq z^2,\end{array} \right\},
\]
defined on a neighborhood of $\{(y,z):(x_o,y,z)\in B_0\}$.
$F$ is called the \textbf{gluing function}.
\end{enumerate}
\end{thm}
\begin{pf} Assume that $\Sigma$ is a smooth
Lorentzian surface in a neighborhood of $B_0$. We split the
proof of the necessary condition in two cases.

\textbf{Case 1:} If both $\alpha$ and $\beta$ do not cross the
axis, each of them is contained in the union of a fundamental region and the axis. We can assume
that $\mathrm{trace}(\alpha)\subseteq
\tilde{\mathscr{R}}^{+}\bigcup\langle\vec{x}\rangle$ and
$\mathrm{trace}(\beta)\subseteq
\tilde{\mathscr{Q}}^{+}\bigcup\langle\vec{x}\rangle$, since the
proof for other cases works similarly.

\vspace{.3cm}

\underline{We prove \textit{LG1}.} 

\vspace{.1cm}

The curve $\alpha$ can be written down as $\alpha(s)=(\alpha_1(s),0,\alpha_3(s))$, and we can assume it is arclength
parametrized. Then, $\alpha_3'(s)^2=\alpha_1'(s)^2 +1>0$. Thus, by using the Inverse Function Theorem, we obtain 
$\rho>0$ such that $\alpha_3:]0,\rho[\longrightarrow\alpha_3(]0,\rho[)$ is a
diffeomorphism. Now, from $\lim_{s\rightarrow 0}\alpha_3(s)=0$ and $\alpha_3>0$, we get that
$\alpha_3(]0,\rho[)=]0,\varrho[$ for certain $\varrho>0$. This provides a smooth function
$f_\alpha :]0,\varrho[\longrightarrow\mathds{R}$, defined by
$f_\alpha(u)=\alpha_1\circ\alpha_3^{-1}(u)$, and so LG1 holds.

\vspace{.3cm}

\underline{We prove \textit{LG2}.} 

\vspace{.1cm}

We can write down $\beta$ as
$\beta(s)=(\beta_1(s),\beta_2(s),0)$, and we can assume this parametrization is arc-length. Then, we consider 
$$X_\beta(s,t):=\xi_t(\beta(s)).$$ 
Our main aim is to show that $\lim_{s\to 0}\beta_2'(s)\neq 0$. Suppose, contrary to our claim, that $\lim_{s\to 0}\beta_2'(s)= 0$.

It should be noticed
that the light-like orbit $\{(x_o,y,y)\in\mathds{L}^3 / y>0\}$ is
contained in $B_0$.
Given a point in that orbit, $q=(x_o,a,a)$ with $a>0$, let
$\gamma_q(\tau)=X_\beta(s_q(\tau),t_q(\tau))$ be a smooth curve with
$\lim_{\tau\to 0}\gamma_q(\tau)=q$. Obviously, this
implies that $\lim_{\tau\to 0}s_q(\tau)=0$. As $\lim_{s\rightarrow 0}\beta_2 (s)=0$, then
\begin{equation}\label{limite-t_q}\lim_{\tau\to
0}\cosh(t_q(\tau))=\lim_{\tau\to
0}\sinh(t_q(\tau))=+\infty.
\end{equation} 
On the other hand, we compute the Riemannian normal to $\Sigma$ along $\gamma_q$, obtaining
$$N_R(\gamma_q(\tau))=\frac{(\beta_2 '(s_q(\tau)), -\beta_1
'(s_q(\tau)) \cosh(t_q(\tau)),\beta_1 '(s_q(\tau))
\sinh(t_q(\tau)))}
{\sqrt{\beta_2'(s_q(\tau))^2+\beta_1'(s_q(\tau))^2(\cosh^2(t_q(\tau))+\sinh^2(t_q(\tau)))}}.$$
Now, by using that $\lim_{s\to 0}\beta_2 '(s)=0$, as well as (\ref{limite-t_q}) and the previous expression, we obtain 
$$N_R(q)=\lim_{\tau\to
0}N_R(s_q(\tau),t_q(\tau))=(0,-\frac{1}{\sqrt{2}},\frac{1}{\sqrt{2}}).$$
This implies that $T_q\Sigma$ is a degenerate plane, which
provides a contradiction. Therefore, we have proven that $\lim_{s\to 0}\beta_2'(s)\neq 0$. Now, we can follow
a similar argument to the one used with $\alpha$, concluding that LG2 holds.

\vspace{.3cm}

\underline{We prove \textit{LG3}.} 

\vspace{.1cm}

We define the following sets
\begin{eqnarray*}
 \Omega &=& \{(y,z)\in\R^2: \vert z^2-y^2\vert<\varrho^2, (x_o,y,z)\in\mathscr{R}^+ \cup \mathscr{Q}^+ \cup B_0\}, \\
\Omega_{\alpha} & = & \{(y,z)\in\R^2: \vert z^2-y^2\vert<\varrho^2, (x_o,y,z)\in\mathscr{R}^+\}, \,\,\mathrm{and}\\
\Omega_{\beta} & = & \{(y,z)\in\R^2: \vert z^2-y^2\vert<\varrho^2,
(x_o,y,z)\in\mathscr{Q}^+\}, 
\end{eqnarray*} 
$\varrho$ being the minimum of $\{\varrho_{\alpha},\varrho_{\beta}\}$.
It is clear that $F_{|\Omega_\alpha}$ and
$F_{|\Omega_\beta}$ are smooth. In order to study the smoothness of $F$ at
the points of $B_0$, we define
$\Sigma^{\prime}=\Sigma_{\alpha_{|]0,\varrho[}}\cup\Sigma_{\beta_{|]0,\varrho[}}\cup
B_0$ and the map $\Pi:\Sigma^{\prime}\longrightarrow \Omega$,
$\Pi(x,y,z)=(y,z)$. $\Pi$ is
smooth, bijective and $\Pi^{-1}(y,z)=(F(y,z),y,z)$. If we prove
that $d\Pi_q$ is bijective $\forall q \in B_0$, the Inverse
Function Theorem gives us the smoothness of $F$ at the points of
$B_0$. It is enough to prove that ${\partial_x}(q)\notin T_q \Sigma'$
$\forall q\in B_0$. To do so, we distinguish three cases:
\begin{itemize}
\item If $q=p$, then
$\{(x_o,y,-y)\in\mathds{L}^3/y\in\mathds{R}\}\subseteq B_0$,
and so, $T_q
\Sigma^{\prime}=\textrm{Span}\{\partial_y(q)+\partial_z(q),\partial_y(q)-\partial_z(q)\}$.

\item If $q\in\{(x_o,y,y)\in\mathds{L}^3/y\neq 0\}$, it is clear
that $\partial_y(q)+\partial_z(q)\in T_q\Sigma^{\prime}$. As $T_q\Sigma^{\prime}$ is a
Lorentzian plane, and $\textrm{Span}\{\partial_x(q),\partial_y(q)+\partial_z(q)\}$ is a
degenerate plane, we get $\partial_x(q)\notin T_q\Sigma^{\prime}$.

\item If $q\in\{(x_o,y,-y)\in\mathds{L}^3/y\neq 0\}$, we proceed similarly to the previous item.
\end{itemize}


\textbf{Case 2:} At least one of the curves $\alpha$ and $\beta$
crosses the axis.

From Case 1, conditions LG1, LG2 and LG3 hold except for the
smoothness of $f_\alpha$ and $f_\beta$ at $u=0$, and the smoothness of $F$ at $(0,0)$ (in the case that
they make sense). 

We will only prove the smoothness of $f_{\alpha}$, since the proof for $f_\beta$ is analogous.

If either $\Sigma_{\alpha}\cap\mathscr{R}^{+}=\emptyset$ or
$\Sigma_{\alpha}\cap\mathscr{R}^{-}=\emptyset$, the proof is
trivial. Consequently, consider that
$\Sigma_{\alpha}\cap\mathscr{R}^{+}\neq\emptyset$ and
$\Sigma_{\alpha}\cap\mathscr{R}^{-}\neq\emptyset$. Then, either
$\Sigma_{\beta}\cap\mathscr{Q}^{+}\neq\emptyset$ or
$\Sigma_{\beta}\cap\mathscr{Q}^{-}\neq\emptyset$. Without loss of
generality, we may assume that
$\Sigma_{\beta}\cap\mathscr{Q}^{+}\neq\emptyset$. Given $a>0$, we
choose the following curves, that are smooth because of Case 1:
$$\gamma_{1}(t) = \Big(F(a-\frac{t}{4a},a+\frac{t}{4a}),a-\frac{t}{4a},a+\frac{t}{4a}\Big)=
\left\{\begin{array}{l}
\left(f_{\alpha}(\sqrt{t}),a-\frac{t}{4a},a+\frac{t}{4a}\right)\ \textrm{if} \  t\geq 0\\
\left(f_{\beta}(\sqrt{-t}),a-\frac{t}{4a},a+\frac{t}{4a}\right)\
\textrm{if}\  t\leq 0
\end{array} \right\},$$
$$\gamma_{2}(t)=
\Big(F(a-\frac{t}{4a},-a-\frac{t}{4a}),a-\frac{t}{4a},-a-\frac{t}{4a}\Big)=\left\{\begin{array}{l}
\left(f_{\alpha}(-\sqrt{t}),a-\frac{t}{4a},-a-\frac{t}{4a}\right) \  \textrm{if} \ t\geq 0 \\
\left(f_{\beta}(\sqrt{-t}),a-\frac{t}{4a},-a-\frac{t}{4a}\right) \
\textrm{if}\   t\leq 0
\end{array} \right\}.$$
Then, we have
$$\lim_{\tiny
\begin{array}{c}t\rightarrow 0\\ t>0\end{array}}\frac{d^n}{dt^n}\left(f_{\alpha}(\sqrt{t})\right)=\lim_{\tiny
\begin{array}{c}t\rightarrow 0\\ t<0\end{array}}\frac{d^n}{d^n
t}\left(f_{\beta}(\sqrt{-t})\right)=\lim_{\tiny
\begin{array}{c}t\rightarrow 0\\ t>0\end{array}}\frac{d^n}{dt^n}\left(f_{\alpha}(-\sqrt{t})\right) \quad \forall n \in
\mathds{N},$$ and consequently
$$\lim_{\tiny
\begin{array}{c}u\rightarrow 0\\ u>0\end{array}}\frac{d^n}{du^n}f_{\alpha}(u)=\lim_{\tiny
\begin{array}{c}u\rightarrow 0\\ u<0\end{array}}\frac{d^n}{du^n}f_{\alpha}(u) \quad \forall n \in \mathds{N}.$$

\vspace{.3cm}

Thanks to Case 1, we can assure that the gluing function is smooth everywhere except at $(0,0)$. The smoothness of $F$ at $(0,0)$ is obtained by applying the Inverse Function Theorem to $\Pi$ at $p$.

\vspace{5mm}

Let us prove the converse. We suppose that LG1, LG2 and LG3 hold. It is
enough to show that $\Sigma$ is a Lorentzian smooth surface in a
neighborhood of $B_0$. Indeed, as the function $F$ can be used to
define a parametrization of $\Sigma$ in a neighborhood of
$B_0$, we only need to exhibit the Lorentzian
character of the surface along $B_0$.

Firstly, we need to prove that $f_\alpha '(0)=f_\beta '(0)=0$. If
$p\in\Sigma$, then the proof is trivial. Otherwise, we have to prove that if any of those two equalities do not hold, then $F$ is not smooth. We know that $B_0$ contains at
least one orbit of the light-like orbital $\mathbf{O}_p$. We
suppose $\{(x_o,y,y)\in\mathds{L}^3/y>0\}\subseteq B_0$ and
we consider $\{(y,y)\,/\,y>0\}$. It is easy to see that the gradient of $F$ is not continuous along $\{(y,y)\,/\,y>0\}$ when any of the equalties $f_\alpha '(0)=f_\beta '(0)=0$ are not true. This is a contradiction because $F$ is a smooth function.

Secondly, 
we prove  the surface along $B_0$ is Lorentzian. If $p\in
B_0$, then $T_p\Sigma=\textrm{Span}\{\lim_{s\to
0}\alpha'(s)=(0,0,1),\lim_{s\to 0}\beta'(s)=(0,1,0)\}$,
which is Lorentzian. Next, we focus on studying the metric along the light-like orbits contained in $B_0$.
If we suppose 
$\{(x_o,y,y)\in\mathds{L}^3/y>0\}\subseteq B_0$, then either
$\Sigma_\alpha\cap\mathscr{R}^+\neq\emptyset$ or
$\Sigma_\beta\cap\mathscr{Q}^+\neq\emptyset$. We assume the first one holds
and we take $(x_o,a,a)$ with $a>0$. Now, choose the curve
$\omega(t)=(F(at,a),at,a)$ for $t\leq 1$. Certainly,
$\omega(1)=(x_o,a,a)$ and so, using that $f_\alpha'(0)=0$, we
compute $\omega'(1)=(-a^2 f_\alpha ''(0),a,0)$.
Then, the vector 
$$\frac{-2}{a(1+a^2(f_\alpha''(0))^2)} \omega'(1) + (0,1,1)$$ is
light-like, it belongs to $T_{(x_o,a,a)}\Sigma$ and it is not proportional to $(0,1,1)$, so $T_{(x_o,a,a)}\Sigma$ is a Lorentzian plane.

The proof for the other light-like orbits is analogous.
\end{pf}

\vspace{4mm}

\noindent\textbf{Remark.} It should be noticed that we have
shown that $f_\alpha '(0)=f_\beta '(0)=0$ when $F$ is smooth. Then, $\alpha$ and
$\beta$ are perpendicular to $\eje$, and so, there cannot exist a
singularity at $p$. Another consequence of this fact is that for each $\mathbf{A}_2$-invariant immersion, the curves $\{\alpha_i:i\in\mathcal{I}\}\cup\{\beta\}$ are smooth.

\subsection{Classification of Lorentzian $\mathbf{A}_2$-invariant surfaces}

As a summary of all the results obtained along the previous
subsections, we exhibit the  classification of Lorentzian
$\mathbf{A}_2$-invariant surfaces.  As a previous step, we need the
following definition.

\begin{definition}\label{general-position} Let
$$\{\beta:\mathrm{J}\rightarrow\mathscr{Q}\}\bigcup\{\alpha_i:\mathrm{J}_{i}\rightarrow
\mathscr{R}\,:\, i\in \mathcal{I}\}$$ be  a countable family of smooth curves such that $\alpha_i$ is time-like and $\mathrm{J}_i\subseteq\mathds{R}$ is an interval $\forall i\in\mathcal{I}$, and $\mathrm{J}$ is either an interval or $\mathds{S}^1$. We will say that these curves are in {\em general position} if they satisfy the following three conditions:
\begin{description}
\item[\textbf{P1.}] $\forall s \in \mathrm{J}$ such that $\beta(s)\in\eje$, there exist $\varepsilon>0$ and  $ i\in\mathcal{I}$ satisfying  
\begin{itemize} 
\item $\beta(s)\in\mathrm{Closure}(\traza{\alpha_i})$, and
\item $\Sigma_{\beta_{|\mathrm{J}_o}}\bigcup\Sigma_{\alpha_i}$ is a smooth surface, where $\mathrm{J}_o=]s_o-\varepsilon,s_o+\varepsilon[\cap\mathrm{J}$.
\end{itemize}
\item[\textbf{P2.}] $\forall i \in \mathcal{I}$, there exist $\varepsilon>0$ and $s_o\in\mathrm{Closure}(\mathrm{J})$ such that 
\begin{itemize}
\item $\lim_{s\to s_o}\beta(s)\in\mathrm{Closure}(\traza{\alpha_i})$, and
\item $\Sigma_{\beta_{|\mathrm{J}_o}}\bigcup\Sigma_{\alpha_i}$ is smooth, where $\mathrm{J}_o=]s_o-\varepsilon,s_o+\varepsilon[\cap\mathrm{J}$.
\end{itemize}
\item[\textbf{P3.}] Using $\mathbf{P1}$ and $\mathbf{P2}$, there can be defined two injective maps,
$$\mathbf{P1}: \{s\in\mathrm{J}:\beta(s)\in\eje\}\longrightarrow \mathcal{I}\ \,\, \mathrm{and}\,\,\,  \mathbf{P2}:\mathcal{I}\longrightarrow\{s_o\in\mathrm{Closure}(\mathrm{J}): \lim_{s\rightarrow s_o}\beta(s)\in\eje\},$$ verifying $\mathbf{P2}(\mathbf{P1}(s))=s$ for all $s\in\mathrm{J}$ such that $\beta(s)\in\eje$.

\end{description}
\end{definition}
\begin{remark}
In the previous definition, when we say $\Sigma_{\beta_{|\mathrm{J}_o}}\bigcup\Sigma_{\alpha_i}$ is smooth, it means that $\Sigma_{\beta_{|\mathrm{J}_o}}\bigcup\Sigma_{\alpha_i}$ joint to an appropriate $\mathbf{A}_2$-invariant subset of $\mathbf{O}_{\beta(s)}$ is smooth.
\end{remark}

\begin{thm} \label{Lorentzianas-Invariantes}
 Let $M$ be a connected surface and $\immersion$ an immersion. Then,
 $(M,\phi^{*}(g))$ is Lorentzian and $\mathbf{A}_2$-invariant if and only if either
 \begin{enumerate}
    \item $\phi(M)$ is a Lorentzian fundamental symmetric surface (described
    in Theorem \ref{Lorentzianas-Invariantes-Fundamentales}), or the union of such surface and one, two, three or four light-like orbits.
    \item $\phi(M)$ is the union of the rotational surfaces generated by a family of curves in general position, $$\{\beta:\mathrm{J}\rightarrow\mathscr{Q}\}\cup\{\alpha_i:\mathrm{J}_{i}\rightarrow
\mathscr{R}\,:\, i\in \mathcal{I}\},$$ and the corresponding
    family of light-like patches, $$\{B_{s_o} : \lim_{s\to s_o}\beta(s)\in\eje\};$$ that is
    $$\phi(M)=\Sigma_{\beta}\bigcup\left(\bigcup_{i\in\mathcal{I}}\Sigma_{\alpha_i}\right)
    \bigcup\left(\bigcup\{B_{s_o} : \lim_{s\to s_o}\beta(s)\in\eje\}\right).$$
 \end{enumerate}
\end{thm}

\subsection{An algorithm to construct $\mathbf{A}_2$-invariant
Lorentzian surfaces not contained in any fundamental region} \label{algoritmo}

To finish the study of $\mathbf{A}_2$-invariant Lorentzian
surfaces, we give an algorithm to construct many examples of this
kind of surfaces, that are not contained in any fundamental
region.
\begin{enumerate}
    \item Given $\delta>0$, choose a smooth function
    $\varphi:\left(-\delta^2,\delta^2\right)\longrightarrow\mathds{R}$.
    \item We consider the functions
    $f_{\alpha}:\mathrm{J}_{\alpha}\subseteq(-\delta,\delta)\longrightarrow\mathds{R}$ and
    $f_{\beta}:\mathrm{J}_{\beta}\subseteq(-\delta,\delta)\longrightarrow\mathds{R}$,
    defined as
\[f_{\alpha}(s)=\varphi(s^2) \quad \textrm{and} \quad    f_{\beta}(s)=\varphi(-s^2),\]
where $\mathrm{J}_{\alpha}$ and $\mathrm{J}_{\beta}$ are
    intervals such that $0$ lies in the closure of both of them and $(f_{\alpha}'(s))^2<1$ for all $s\in\mathrm{J}_{\alpha}$.
    \item We define the following curves,
    $\alpha:\mathrm{J}_{\alpha}\longrightarrow\mathscr{R}$ and
    $\beta:\mathrm{J}_{\beta}\longrightarrow\mathscr{Q}$, given by \[\alpha(s)=(f_{\alpha}(s),0,s) \quad \textrm{and} \quad
\beta(s)=(f_{\beta}(s),s,0).\]

\item Choose $B_0\subset\{(\varphi(0),y,z)\in\Ltres : y^2=z^2\}$  to be $\mathbf{A}_2$-invariant and such that $\Sigma_{\alpha}\cup\Sigma_{\beta}\cup B_0$ is a topological surface. 

\item The surfaces $\Sigma_{\alpha}$ and $\Sigma_{\beta}$, generated by $\alpha$ and $\beta$ respectively, glue smoothly and  $\Sigma=\Sigma_{\alpha}\bigcup\Sigma_{\beta}\bigcup B_0$ is an $\mathbf{A}_2$-invariant Lorentzian surface. The gluing function (see Theorem \ref{Pegado}) is given by
\[F(y,z)=\varphi(z^2-y^2).\]
\end{enumerate}

In addition, if we start with two functions, $f_{\alpha}$ and
$f_{\beta}$, such that they are analytic in $0$,
$f_{\alpha}(0)=f_{\beta}(0)$ and the corresponding graphs in
$\mathscr{R}$ and $\mathscr{Q}$ generate Lorentzian surfaces that
glue smoothly with Lorentzian metric along the union (which means that the gluing function $F$ is
smooth), then we can show the existence of a smooth function,
$\varphi$, that allows us to write those graphs as
\[\alpha(s)=(\varphi(s^2),0,s) \quad \textrm{and} \quad \beta(s)=(\varphi(-s^2),s,0).\]
This result constitutes a kind of converse when starting from analytic
data.

The following pictures illustrate the last four subsections. Picture
A shows a surface touching the axis twice, and crossing all
fundamental regions in both cases. Picture B shows a surface
touching the axis at three points, crossing four fundamental
regions around the first point, three regions around the second point and
just one region around the final point.

\begin{center}
\begin{tabular}{|c|c|} \hline
\includegraphics[scale=0.8]{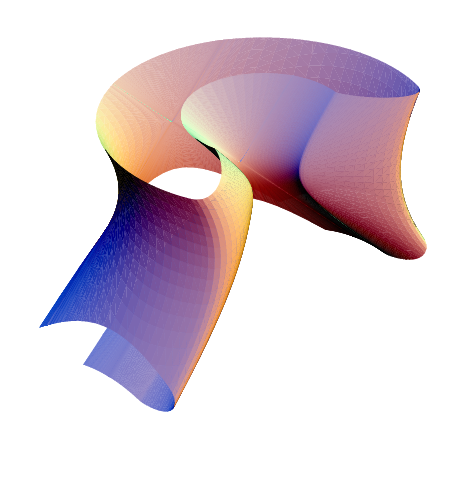} & \includegraphics[scale=0.85]{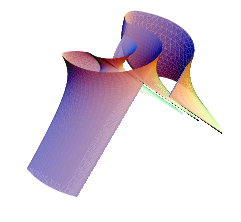}
\\ \hline
Picture A & Picture B \\ \hline
\end{tabular}
\end{center}

\section{First variation of the Willmore functional in a semi-Riemannian manifold} \label{variacion-will}

In contrast with the pure rotational case, i.~e., the one associated
with the group $\mathbf{A}_1$, now, we will use a direct
variational approach to study the case of space-like axis. This
means we will avoid the principle of symmetric criticality. Thus,
it is necessary to obtain the Euler-Lagrange equations associated
with the problem
$\left(\mathbf{I}_{\Gamma}^{\varepsilon}(M,\mathds{L}^3);\mathfrak{S}\right)$
or equivalently
$\left(\mathbf{I}_{\Gamma}^{\varepsilon}(M,\mathds{L}^3);\mathfrak{W}\right)$.
These equations were computed in \cite{Weiner}, when the target
space was a Riemannian three-space with constant curvature.
However, now we have semi-Riemannian target spaces, namely
Lorentzian three-spaces. On the other hand, the constancy of the
curvature is not enough for our purposes. In fact, we will need to
make some suitable conformal changes, in the Lorentz-Minkowski
metric, which, obviously, will not preserve the constancy of the
curvature. Consequently, the variational setting will be as
general as possible and later, computations will be particularized
to our purposes. To start with, we introduce some preliminaries
following the notation of \cite{Weiner}.

Let $\mathsf{M}$ be a compact orientable smooth surface with
 boundary (maybe empty) and $(\bar{\mathsf{M}},\bar{\mathsf{g}})$ a
3-dimensional semi-Riemannian manifold. Let $\phi:\mathsf{M}
\rightarrow \bar{\mathsf{M}}$ be a non-degenerate immersion. Only in
this section, $\mathbf{I}_{\phi(\partial
\mathsf{M})}(\mathsf{M},\bar{\mathsf{M}})$ will denote the space of
non-degenerate immersions that fix the boundary, $\phi(\partial
\mathsf{M})$, without further conditions on the normal field along
the common boundary.

A variation of $\phi$ in $\mathbf{I}_{\phi(\partial
\mathsf{M})}(\mathsf{M},\bar{\mathsf{M}})$ is nothing but a smooth
map,
$\Phi:\mathsf{M}\times(-\delta,\delta)\longrightarrow\bar{\mathsf{M}}$
satisfying the following conditions:
\begin{enumerate}
    \item For each $v\in(-\delta,\delta)$, the map
    $\phi_v:\mathsf{M}\longrightarrow\bar{\mathsf{M}}$, defined by
    $\phi_v(m)=\Phi(m,v)$, belongs to $\mathbf{I}_{\phi(\partial
\mathsf{M})}(\mathsf{M},\bar{\mathsf{M}})$, and
    \item $\phi_0=\phi$.
\end{enumerate}
It should be noticed that $\phi_v^{*}(\bar{\mathsf{g}})$ is
non-degenerate for any
    $v\in(-\delta,\delta)$, and  $\Phi(m,v)=\phi(m)$, for any $m\in\partial \mathsf{M}$.
Now, we can use all the paraphernalia of geometrical objects along a
map. In particular, we can talk about vector fields along $\Phi$, or in
other words, cross sections of the induced vector bundle
$\Phi^{*}(\mathsf{T}\bar{\mathsf{M}})$ over
$\mathsf{M}\times(-\delta,\delta)$. Thus, we can define the
following vector field along $\Phi$, 
\[\mathbf{V}(m,v)=\Phi_{*}\left(\frac{\partial}{\partial
v}(m,v)\right).  \]
In particular, it holds $\mathbf{V}(m,v)=0$, $\forall m\in\partial \mathsf{M}$.
This, when restricted to $v=0$, provides a vector field along $\phi$ which
vanishes along $\partial \mathsf{M}$,  called the variational
vector field
\[\mathbf{V}(m)=\mathbf{V}(m,0)=\Phi_{*}\left(\frac{\partial}{\partial
v}(m,0)\right).\]
Therefore, the tangent space
$\mathsf{T}_{\phi}\left(\mathbf{I}_{\phi(\partial
\mathsf{M})}(\mathsf{M},\bar{\mathsf{M}})\right)$ is made up of
those vector fields along $\phi$ that vanish along $\partial
\mathsf{M}$.

We consider the Willmore variational problem which is associated
with the functional $\mathfrak{W}:\mathbf{I}_{\phi(\partial
\mathsf{M})}(\mathsf{M},\bar{\mathsf{M}})\longrightarrow\mathds{R}$
defined by
$$
     \mathfrak{W}(\psi)=\int_{\mathsf{M}}\,\left(H_{\psi}^2+\mathsf{R}_{\psi}\right)\,dA_{\psi}+\int_{\partial \mathsf{M}}\,\kappa_{\psi}\,ds,
$$
where $H_{\psi}$ denotes the mean curvature function of
$(\mathsf{M},\psi)$, $\mathsf{R}_{\psi}$ is the sectional curvature
of the target space, $(\bar{\mathsf{M}},\bar{\mathsf{g}})$,
restricted to the tangent plane $d\psi(\mathsf{T} \mathsf{M})$ and
$\kappa_{\psi}$ is the geodesic curvature of $\partial \mathsf{M}$
in $(\mathsf{M},\psi^*(\bar{\mathsf{g}}))$.
Now, we wish to determine the sufficient and necessary conditions for $\phi$
to be a critical point of the above
functional, in other words, a Willmore surface of the conformal
space $(\bar{\mathsf{M}},[\bar{\mathsf{g}}])$. Therefore, we need to
compute the differential of $\mathfrak{W}$ at $\phi$, i.~e., 
$\delta\mathfrak{W}(\phi):\mathsf{T}_{\phi}\left(\mathbf{I}_{\phi(\partial \mathsf{M})}(\mathsf{M},\bar{\mathsf{M}})\right)\longrightarrow\mathds{R}$.
Given $\mathbf{V}\in\mathsf{T}_{\phi}\left(\mathbf{I}_{\phi(\partial
\mathsf{M})}(\mathsf{M},\bar{\mathsf{M}})\right)$, consider a
variation
$\Phi:\mathsf{M}\times(-\delta,\delta)\longrightarrow\bar{\mathsf{M}}$
with
$\mathbf{V}(m)=\mathbf{V}(m,0)=\Phi_{*}\left(\frac{\partial}{\partial
v}(m,0)\right)$, $\forall m\in \mathsf{M}$. To simplify the
notation, we put
$\mathsf{M}_v=(\mathsf{M},\phi_v^*(\bar{\mathsf{g}}))$,
$H_v=H_{\phi_v}$, $\mathsf{R}_v=\mathsf{R}_{\phi_v}$,
$\kappa_v=\kappa_{\phi_v}$, $dA_v=dA_{\phi_v}$,
 and then
\begin{equation} \label{diferencial-Willmore}
\delta\mathfrak{W}(\phi)[\mathbf{V}]=\left\{\frac{\partial}{\partial
v}\left[\int_{\mathsf{M}_v}\left(H_v^2+\mathsf{R}_v\right)dA_v+\int_{\partial
\mathsf{M}}\kappa_v\, ds\right]\right\}_{v=0}.
\end{equation}
We will compute this step by step. First, we control the action on
the boundary. To do so, we recall that we are using variations
that fix the boundary, $\phi(\partial \mathsf{M})$. Let
$\{\nu,T\}$ be a unitary positively oriented frame field on
$\phi(\partial \mathsf{M})$, where $T$ is tangent to
$\phi(\partial \mathsf{M})$ and $\nu$ is the outward normal to
$\phi(\partial \mathsf{M})$. 
Since $T$ does not depend on $v$, $\{\nu^{v},T\}$ is the
orientation of $\phi(\partial \mathsf{M})$ in $\phi_{v}(M)$, for
each $v$. Moreover, the principal curvature vector field,
$\eta=\bar{\nabla}_T T$, of $\phi(\partial \mathsf{M})$ in
$(\bar{\mathsf{M}},\bar{\mathsf{g}})$, does not depend on $v$, and
so
\begin{equation} \label{variacion-curvatura-geodesica}\left\{\frac{\partial}{\partial v}\int_{\partial \mathsf{M}}\kappa_v\,
ds\right\}_{v=0}=-\int_{\partial
\mathsf{M}}\bar{\mathsf{g}}(\eta,\bar{\nabla}_{\mathbf{V}} \nu^v)ds=
-\int_{\partial
\mathsf{M}}\bar{\mathsf{g}}\left(\eta^{\perp},D_{\nu}\mathbf{V}^{\perp}\right)ds,\end{equation}
where $\perp$ indicates normal component and $D$ the normal connection
of $(\mathsf{M},\phi)$ in $(\bar{\mathsf{M}},\bar{\mathsf{g}})$. To
obtain the second equality we have used an argument similar to that
used in the Riemannian case, \cite{Weiner}.

\vspace{2mm}

\noindent \textbf{Remark.} Regarding
(\ref{variacion-curvatura-geodesica}), it should be noticed that
under the boundary conditions we are considering in this paper
(i.~e., space of surfaces immersed in $\mathds{L}^3$, with the same
causal character, the same boundary and the same Gauss map along
the common boundary), $D_{\nu}\mathbf{V}^{\perp}=0$ and so
\[\left\{\frac{\partial}{\partial v}\int_{\partial \mathsf{M}}\kappa_v\,
ds\right\}_{v=0}=0,
\]
which is not surprising, because the total curvature of the boundary is a
constant under these boundary conditions.

To obtain the variation of the two-dimensional integral appearing
in (\ref{diferencial-Willmore}), we need some formulae which can
be obtained by using standard variational arguments and that we
collect in the following

\begin{lem} The following statements hold
\begin{enumerate}
\item Let $\mathbf{H}(m,v)$ be the vector field along the variation
$\Phi$ that measures the mean curvature vector field of
$(\mathsf{M},\phi_v)$ at $m\in \mathsf{M}$, then
\[\left\{D_{\frac{\partial}{\partial v}}\mathbf{H}\right\}_{v=0}=\frac{1}{2}\left[\triangle \mathbf{V}^{\perp}+\tilde{A}(\mathbf{V}^{\perp})
+\varepsilon\textrm{Ric}(N_{\phi},N_{\phi})\mathbf{V}^{\perp}\right]+D_{\mathbf{V}^{\top}}\mathbf{H},\]
where $\triangle$ is the Laplacian relative to the normal
connection, $D$, $\tilde{A}$ is the Simons' operator (see
\cite{Simons}), $\textrm{Ric}$ is the Ricci tensor of
$(\bar{\mathsf{M}},\bar{\mathsf{g}})$ and
$\varepsilon=\bar{\mathsf{g}}(N_{\phi},N_{\phi})$.
\item The variation of the area element is given by the following
formula
\[\left\{\frac{d}{dv}\left(dA_v\right)\right\}_{v=0}=-2\bar{\mathsf{g}}(\mathbf{H},\mathbf{V})\,dA+d\theta,\]
where $dA=dA_0$ and $\theta$ is the one-form defined by
$\theta(Z)=dA(\mathbf{V}^\top,Z)$. \end{enumerate}
\end{lem}

\vspace{2mm}

\noindent \textbf{Remark.} The proof of $1$ can be found in
\cite{Weiner}. The proof of $2$ is almost the same as in the
Riemannian case.

\vspace{2mm}

\noindent Next, we use the above lemma joint the fact that
$H_v^2=\varepsilon\bar{\mathsf{g}}(\mathbf{H},\mathbf{H})$ to
obtain
\begin{eqnarray*}
&\left\{ \displaystyle \frac{\partial}{\partial
v}\left[\left(H_v^2+\mathsf{R}_v\right)dA_v\right]\right\}_{v=0} =
\left[\varepsilon\bar{\mathsf{g}}(\triangle\mathbf{V}^{\bot},\mathbf{H})+\mathbf{V}^{\top}(H^2)+\left(\frac{\partial
\mathsf{R}_v}{\partial v}\right)_{v=0}\right]dA &  \\
& +\bar{\mathsf{g}}\left(\varepsilon\tilde{A}(\mathbf{H})+\textrm{Ric}(N_{\phi},N_{\phi})\mathbf{H}-2(H^2+\mathsf{R})\mathbf{H},\mathbf{V}^{\bot}\right)dA +(H^2+\mathsf{R})\,d\theta. &
\end{eqnarray*}
On the other hand,
$(H^2+\mathsf{R})\,d\theta=d\left((H^2+\mathsf{R})\,\theta\right)-\left(\mathbf{V}^{\top}(H^2+\mathsf{R})\right)dA$.
Since $\theta$ vanishes on $\partial \mathsf{M}$, we have
\[\int_{\mathsf{M}}(H^2+\mathsf{R})\,d\theta=-\int_{\mathsf{M}}\left(\mathbf{V}^{\top}(H^2+\mathsf{R})\right)dA.\]
It is easy to see that Proposition $1.2.$ in \cite{Weiner} remains
true in a semi-Riemannian setting. Then, we make use of it to have
\begin{eqnarray*}\left\{\frac{\partial}{\partial
v}\int_{\mathsf{M}_v}\left(H_v^2+\mathsf{R}_v\right)dA_v\right\}_{v=0}&=&\int_{\mathsf{M}}\left[\bar{\mathsf{g}}(\mathfrak{R}(\mathbf{H}),\mathbf{V}^{\bot})+
\mathbf{V}^{\bot}(\mathsf{R}^\mathbf{V})\right]dA
\\ &+& \varepsilon\int_{\partial
\mathsf{M}}\bar{\mathsf{g}}(\mathbf{H},D_{\nu}\mathbf{V}^{\bot})ds,
\end{eqnarray*} where
$\mathfrak{R}=\varepsilon(\triangle+\tilde{A})+(\textrm{Ric}(N_{\phi},N_{\phi})-2(H^2+\mathsf{R}))\,\mathbf{I}$
is a kind of Schr\"{o}dinger operator, $\mathbf{I}$ is the identity map 
and $\mathsf{R}^\mathbf{V}(\Phi(m,v))=\mathsf{R}_v(m)$. Finally, we combine
this formula with (\ref{variacion-curvatura-geodesica}) to get
$$
\delta\mathfrak{W}(\phi)[\mathbf{V}]=
\int_{\mathsf{M}}\left[\bar{\mathsf{g}}(\mathfrak{R}(\mathbf{H})+(\nabla
\mathsf{R}^\mathbf{V})^{\bot},\mathbf{V}^{\bot})\right]dA + \int_{\partial
\mathsf{M}}\bar{\mathsf{g}}(\varepsilon\mathbf{H}-\eta^{\bot},D_{\nu}\mathbf{V}^{\bot})ds,
$$
where $(\nabla \mathsf{R}^\mathbf{V})^{\bot}=\varepsilon
N_{\phi}(\mathsf{R}^\mathbf{V})N_{\phi}$ is the normal component of the
gradient of $\mathsf{R}^\mathbf{V}$.

These computations can be summarized in the following result which
gives the first variation of the Willmore functional in a
semi-Riemannian manifold, $(\bar{\mathsf{M}},[\bar{\mathsf{g}}])$

\begin{thm} \label{Willmore-Surfaces}
In the previous setting, $(\mathsf{M},\phi)$
is a Willmore surface in $(\bar{\mathsf{M}},[\bar{\mathsf{g}}])$ with boundary date $\phi(M)$, if
and only if
$$\int_{\mathsf{M}}\left[\bar{\mathsf{g}}(\mathfrak{R}(\mathbf{H})+\varepsilon
N_{\phi}(\mathsf{R}^\mathbf{V})N_{\phi},\mathbf{V}^{\bot})\right]dA +
 \int_{\partial
\mathsf{M}}\bar{\mathsf{g}}(\varepsilon\mathbf{H}-\eta^{\bot},D_{\nu}\mathbf{V}^{\bot})ds=0,
$$
for any $\mathbf{V}\in \mathsf{T}_{\phi}\left(\mathbf{I}_{\phi(\partial
\mathsf{M})}(\mathsf{M},\bar{\mathsf{M}})\right)$.
\end{thm}

From now on, we assume the boundary conditions we are considering
along this paper. Namely, let
$\Gamma=\{\gamma_1,\gamma_2,\dots,\gamma_n\}$ be a finite set of
non-null regular curves in $\bar{\mathsf{M}}$ with
$\gamma_i\bigcap\gamma_j=\emptyset$, if $i\neq j$ and choose $N_o$
to be a unitary vector field along $\Gamma$ which is orthogonal to
$\Gamma$ and has constant causal character, $\varepsilon$, on the
whole $\Gamma$. Then, we consider the space of immersions,
$\mathbf{I}_{\Gamma}^{\varepsilon}(\mathsf{M},\bar{\mathsf{M}})$,
made up of those immersions,
$\phi:\mathsf{M}\longrightarrow\bar{\mathsf{M}}$ satisfying
\begin{center}
    $\phi(\partial \mathsf{M})=\Gamma$, \quad $N_{\phi}=N_o$ along $\partial \mathsf{M}$ \quad and \quad $\bar{\mathsf{g}}(N_{\phi},N_{\phi})=\varepsilon$.
\end{center}
In this case (see the above remark), $D_{\nu}(\mathbf{V})^{\perp}=0$
and therefore, the boundary term vanishes. Consequently, the Willmore
surfaces with prescribed Gauss map along the common boundary in
$(\bar{\mathsf{M}},[\bar{\mathsf{g}}])$ are characterized by the equation
\begin{equation} \label{Willmore-surfaces-prescribed-Gauss-map}
\int_{\mathsf{M}}\left[\bar{\mathsf{g}}(\mathfrak{R}(\mathbf{H})+\varepsilon
N_{\phi}(\mathsf{R}^\mathbf{V})N_{\phi},\mathbf{V}^{\bot})\right]dA=0, \quad
\forall\, \mathbf{V}\in
\mathsf{T}_{\phi}\left(\mathbf{I}_{\Gamma}^{\varepsilon}(\mathsf{M},\bar{\mathsf{M}})\right).
\end{equation}
An easy computation shows  $\triangle\mathbf{H}=(\triangle
H)N_{\phi}$ and $\tilde{A}(\mathbf{H})=\varepsilon H\|dN_{\phi}\|^2
N_{\phi}$, allowing us to reduce
(\ref{Willmore-surfaces-prescribed-Gauss-map}) to
\begin{equation}
\label{Scalar-Willmore-surfaces-prescribed-Gauss-map}
\int_{\mathsf{M}}\left[\varepsilon\triangle
H+H\left(\|dN_{\phi}\|^2+\textrm{Ric}(N_{\phi},N_{\phi})-2(H^2+\mathsf{R})\right)+\varepsilon
N_{\phi}(\mathsf{R}^\mathbf{V})\right]\bar{\mathsf{g}}(N_{\phi},\mathbf{V}^{\bot})dA=0,
\end{equation}
for any $\mathbf{V}\in  \mathsf{T}_{\phi}\left(\mathbf{I}_{\Gamma}^{\varepsilon}(\mathsf{M},\bar{\mathsf{M}})\right)$.

\section{$\mathbf{A}_2$-invariant Riemannian solutions} \label{Ados-R-solutiones}

$\mathbf{A}_2$-invariant Riemannian surfaces in $\mathds{L}^3$
were classified in Theorem \ref{Riemannianas-Invariantes}. 
Each surface of this type is generated by a curve, $\alpha$, immersed in either 
$\tilde{\mathscr{R}}^{+}$ or
$\tilde{\mathscr{R}}^{-}$. This curve evolves according to the motions of $\mathbf{A}_2$ to produce the symmetric surface
$$ \Sigma_{\alpha}  =\{\xi_t(\alpha(s))\, : \, s\in I, \, t\in\mathds{R}\},$$ lying in either $\mathscr{R}^+$ or $\mathscr{R}^-$
respectively.

At this time, our aim consists of finding which curves generate surfaces $\Sigma_{\alpha}$ which are Riemannian solutions of the $O(2,1)$ Nonlinear Sigma
Model.
Since
$\left(\mathbf{I}_{\partial\Sigma_{\alpha}}^{-1}(M,\mathds{L}^3);\mathfrak{S}\right)$
is equivalent to
$\left(\mathbf{I}_{\partial\Sigma_{\alpha}}^{-1}(M,\mathds{L}^3);\mathfrak{W}\right)$,
the above problem is reduced to finding which curves $\alpha$ generate Riemannian Willmore surfaces with prescribed
time-like Gauss map along the boundary, $\partial\Sigma_{\alpha}$. The solution is made explicit in the following result. For a better understanding, we recall Lemma \ref{warped-product}, where we showed that the Lorentz-Minkowski metric,
$g$, on both $\mathscr{R}^{+}$ and $\mathscr{R}^{-}$ is conformal to
a semi-Riemannian product of a de Sitter plane and a space-like
Lorentzian circle.

\begin{thm} \label{Riemannian-solutions} Let $\alpha$ be a space-like curve immersed in either $\tilde{\mathscr{R}}^{+}$ or $\tilde{\mathscr{R}}^{-}$. Then,
$\Sigma_{\alpha}$ is a Riemannian solution of the $O(2,1)$
Nonlinear Sigma Model, that is, a Riemannian Willmore surface with
prescribed Gauss map along the common boundary in
$(\mathds{L}^3,g)$, if and only if $\alpha$ is a free elastica in
the de Sitter plane
$\left(\tilde{\mathscr{R}}^{+},\frac{1}{f_{+}^2}g\right)$ or in
the de Sitter plane
$\left(\tilde{\mathscr{R}}^{-},\frac{1}{f_{-}^2}g\right)$, 
respectively.
\end{thm}

\begin{pf}  We consider the case where $\alpha$ is immersed in
$\tilde{\mathscr{R}}^{+}$, since the  other case is analogous. The
proof is obtained from a chain of three steps.

\textit{Step 1: We  make use of the conformal invariance of the Willmore
functional}. In this setting, the suitable boundary conditions for
the problem are $\Gamma=\partial\Sigma_{\alpha}$  and
$N_o=N_{\Sigma_{\alpha}}$. Even more, as $\Sigma_{\alpha}$ is the
image of an immersion from $[a_1,a_2]\times\mathds{R}$, we deduce
that $M=[a_1,a_2]\times\mathds{R}$. Notice that
$\mathbf{I}_{\partial\Sigma_{\alpha}}^{-1}(M,\mathscr{R}^{+})$ is
an open set of
$\mathbf{I}_{\partial\Sigma_{\alpha}}^{-1}(M,\mathds{L}^3)$. Using
this fact, as well as the conformal invariance of the Willmore
functional, it is easy to see that $\Sigma_{\alpha}$ is a
Riemannian Willmore surface with prescribed Gauss map along the
common boundary in $(\mathds{L}^3,g)$, if and only if
$\Sigma_{\alpha}$ is a solution of
$(\mathbf{I}_{\partial\Sigma_{\alpha}}^{-1}(M,
\mathscr{R}^{+});\overline{\mathfrak{W}})$, where
$\overline{\mathfrak{W}}$ is the Willmore functional of
$\left(\mathscr{R}^{+},\frac{1}{f_{+}^2}g\right)$.

From now on, we denote with a bar all the elements in
$\left(\mathscr{R}^{+},\frac{1}{f_{+}^2}g\right)$. Also, to
simplify the notation, we put
$\bar{N}_{\alpha}=\bar{N}_{\Sigma_{\alpha}}$ and
$\bar{H}_{\alpha}=\bar{H}_{\Sigma_{\alpha}}$. Denote by
$\bar{\mathsf{R}}_{\alpha}$ the sectional curvature of
$\left(\mathscr{R}^{+},\frac{1}{f_{+}^2}g\right)$ along
$\Sigma_{\alpha}$ and by $\bar{K}_{\alpha}$ the Gaussian curvature
of $\Sigma_{\alpha}$ with the induced metric from
$\left(\mathscr{R}^{+},\frac{1}{f_{+}^2}g\right)$.

\textit{Step 2: We compute the term
$\bar{N}_{\alpha}(\mathsf{\bar{R}}^\mathbf{V})$}. The surface
$\Sigma_{\alpha}$ is a critical point of the problem
$(\mathbf{I}_{\partial\Sigma_{\alpha}}^{-1}(M,
\mathscr{R}^{+}));\overline{\mathfrak{W}})$ if and only if,
(\ref{Scalar-Willmore-surfaces-prescribed-Gauss-map}) holds on
every non-null polygon contained in $\Sigma_{\alpha}$, for
$\varepsilon=-1$. However, it will be interesting to obtain a
characterization of the critical points of
$(\mathbf{I}_{\partial\Sigma_{\alpha}}^{-1}(M,
\mathscr{R}^{+});\overline{\mathfrak{W}})$ as the solution of an
equation involving terms that depend only on $\Sigma_{\alpha}$. In
this sense, one needs to manipulate the term
$\bar{N}_{\alpha}(\mathsf{\bar{R}}^\mathbf{V})$ because, a priori, it is
the only one that depends on $\mathbf{V}$. Pick a  point
$p\in\Sigma_{\alpha}$, and compute $\mathsf{\bar{R}}^\mathbf{V}$ along a
curve, $\gamma$, with $\gamma(0)=p$ and
$\gamma^{\prime}(0)=\bar{N}_{\alpha}(p)$. Let
$\phi_{\alpha}:I\times\mathds{R}\longrightarrow\left(\mathscr{R}^{+},\frac{1}{f_{+}^2}g\right)$
be the parametrization of $\Sigma_{\alpha}$ defined as
$\phi_{\alpha}(s,t)=\xi_t(\alpha(s))$. Write
$\alpha(s)=(\alpha_{1}(s),0,\alpha_{3}(s))$ so that
$\frac{1}{f_{+}^2}g(\alpha',\alpha')=1$, then a unitary normal vector
to $\Sigma_{\alpha}$ can be computed to be
$$\bar{N}_{\alpha}(\phi_{\alpha}(s,t))=\bar{N}_{\alpha}(s,t)=\xi_t((\alpha_{3}^{\prime}(s),0,\alpha_{1}^{\prime}(s))=(\alpha_3'(s),\alpha_1'(s)\sinh
t,\alpha_1'(s)\cosh t).$$ Given
$\mathbf{V}\in\mathsf{T}_{\phi_{\alpha}}\left(\mathbf{I}_{\partial\Sigma_{\alpha}}^{-1}(M,
\mathscr{R}^{+}))\right)$ with compact support (remember we are
working with non null polygons in $\Sigma_{\alpha}$), an
associated variation of $\phi_{\alpha}$ is just
$$\Phi:(-\delta,\delta)\times I\times\mathds{R}\longrightarrow\left(\mathscr{R}^{+},\frac{1}{f_{+}^2}g\right), \quad
\Phi(v,s,t)=\phi_{\alpha}(s,t)+v \mathbf{V}(\phi_{\alpha}(s,t)).$$
Without loss of generality, we can assume $\mathbf{V}^{\top}=0$.
In this case, there exists a function
$f:I\times\mathds{R}\longrightarrow\mathds{R}$ with compact
support, such that $\mathbf{V}=f\bar{N}_{\alpha}$, and so
$\Phi(v,s,t)=\phi_{\alpha}(s,t)+v f(s,t)\bar{N}_{\alpha}(s,t)$.
Given a point $\Phi(v_{o},s_{o},t_{o})$, then
$\mathsf{\bar{R}}^\mathbf{V}(\Phi(v_{o},s_{o},t_{o}))$ is the sectional
curvature of the plane generated by
$\{(\partial_{s}\Phi)(v_{o},s_{o},t_{o}),(\partial_{t}\Phi)(v_{o},s_{o},t_{o})\}$.
Recall that $\left(\mathscr{R}^{+},\frac{1}{f_{+}^2}g\right) =
\left(\tilde{\mathscr{R}}^{+},\frac{1}{f_{+}^2}g\right)\times
(\mathrm{H}^{+},dt^2)$ and denote by
$\Pi^1:\left(\mathscr{R}^{+},\frac{1}{f_{+}^2}g\right)\longrightarrow\left(\tilde{\mathscr{R}}^{+},\frac{1}{f_{+}^2}g\right)$
and
$\Pi^2:\left(\mathscr{R}^{+},\frac{1}{f_{+}^2}g\right)\longrightarrow(\mathrm{H}^{+},dt^2)$
the canonical projections. It is clear that
$\bar{\mathcal{R}}(\Phi_s,\Phi_t,\Phi_t,\Phi_s)=\tilde{\mathcal{R}}(E_s,E_t,E_t,E_s)$,
where $\bar{\mathcal{R}}$ is the curvature tensor of
$\left(\mathscr{R}^{+},\frac{1}{f_{+}^2}g\right)$ and
$\tilde{\mathcal{R}}$ stands for the curvature tensor of
$\left(\tilde{\mathscr{R}}^{+},\frac{1}{f_{+}^2}g\right)$,
$E_s=d\Pi^{1}(\partial_{s}\Phi)$ and
$E_t=d\Pi^{1}(\partial_{t}\Phi)$. Therefore,
$$\mathsf{\bar{R}}^\mathbf{V}(\Phi(v_{o},s_{o},t_{o}))=\frac{\tilde{\mathcal{R}}(E_s,E_t,E_t,E_s)}{\frac{1}{f_{+}^2}g(\partial_{s}\Phi,\partial_{s}\Phi)\frac{1}{f_{+}^2}g(\partial_{t}\Phi,\partial_{t}\Phi)-
(\frac{1}{f_{+}^2}g(\partial_{s}\Phi,\partial_{t}\Phi))^2}.$$
Observe that the normal vector at $\phi_{\alpha}(s_{o},t_{o})$ is
$\bar{N}_{\alpha}(s_{o},t_{o})=\gamma^{\prime}(0)$, where
$$\gamma(\tau)=\phi_{\alpha}(s_{o},t_{o})+\tau\bar{N}_{\alpha}(s_{o},t_{o})=\Phi\left(\frac{\tau}{f(s_{o},t_{o})},s_{o},t_{o}\right).$$
To compute the value of $\mathsf{\bar{R}}^\mathbf{V}$ along $\gamma$, we
have
$$\tilde{\mathcal{R}}(E_s,E_t,E_t,E_s) =
\tau^2(\frac{\partial_{t}f}{f})^2[\tilde{\mathcal{R}}(U_o,V_1,V_1,U_o)+2\tau\tilde{\mathcal{R}}(U_o,V_1,V_1,U_1)+
\tau^2\tilde{\mathcal{R}}(U_1,V_1,V_1,U_1)],$$ being
$U_o=(\alpha_1^{\prime},0,\alpha_3^{\prime})$,
$U_1=(\frac{\partial_s
f}{f}\alpha_3^{\prime}+\alpha_3^{\prime\prime},0,\frac{\partial_s
f}{f}\alpha_1^{\prime}+\alpha_1^{\prime\prime})$ and
$V_1=(\alpha_3^{\prime},0,\alpha_1^{\prime})$. Also
$$\frac{1}{f_{+}^2}g(\partial_{s}\Phi,\partial_{s}\Phi)\frac{1}{f_{+}^2}g(\partial_{t}\Phi,\partial_{t}\Phi)-
\left(\frac{1}{f_{+}^2}g(\partial_{s}\Phi,\partial_{t}\Phi)\right)^2
= \frac{\alpha_{3}^4
+b_1\tau+b_2\tau^2+b_3\tau^3+b_4\tau^4}{(\alpha_3
+\tau\alpha_{1}^{\prime})^4},$$ where $b_1$, $b_2$, $b_3$ and $b_4$
are functions that do not depend on $\tau$.
Then,
$$\mathsf{\bar{R}}^\mathbf{V}(\Phi(\frac{\tau}{f(s_o,t_o)},s_{o},t_{o}))=$$
$$=\tau^2\left(\frac{\partial_{t}f}{f}\right)^2\frac{(\alpha_3
+\tau\alpha_{1}^{\prime})^4
\left[\tilde{\mathcal{R}}(U_o,V_1,V_1,U_o)+2\tau\tilde{\mathcal{R}}(U_o,V_1,V_1,U_1)+\tau^2\tilde{\mathcal{R}}(U_1,V_1,V_1,U_1)\right]}{\alpha_{3}^4
+b_1\tau+b_2\tau^2+b_3\tau^3+b_4\tau^4}.$$ Since $\alpha_3>0$, we
obtain
$$\bar{N}_{\alpha}(\mathsf{\bar{R}}^\mathbf{V})(\phi(s_o,t_o))=\frac{d}{d\tau}_{|\tau=0}\left(\mathsf{\bar{R}}^\mathbf{V}\left(\Phi\left(\frac{\tau}{f(s_o,t_o)},s_{o},t_{o}\right)\right)\right)=0.$$

\textit{Step 3: The solutions come from clamped elasticae in the
de Sitter plane
$\left(\tilde{\mathscr{R}}^{+},\frac{1}{f_{+}^2}g\right)$. } The
computation we did in the last step allows one to characterize the
solutions, $\Sigma_{\alpha}$, of
$(\mathbf{I}_{\partial\Sigma_{\alpha}}^{-1}([a_1,a_2]\times\mathds{R},
\mathscr{R}^{+});\overline{\mathfrak{W}})$ as the solutions of the
following Euler-Lagrange equation

\begin{equation}\label{equation-Riemannian-critical-point}-\triangle
\bar{H}_{\alpha}+\bar{H}_{\alpha}\left(\|d\bar{N}_{\alpha}\|^2+\textrm{Ric}(\bar{N}_{\alpha},\bar{N}_{\alpha})-2(\bar{H}_{\alpha}^2+\bar{\mathsf{R}}_{\alpha})\right)=0
\quad \textrm{ in } \Sigma_{\alpha}.
\end{equation}
Using
$\|d\bar{N}_{\alpha}\|^2=4\bar{H}_{\alpha}^2-2\det(-d\bar{N}_{\alpha})$
and $\bar{K}_{\alpha}=\bar{\mathsf{R}}_{\alpha}-\det(-d\bar{N}_{\alpha})$,
equation (\ref{equation-Riemannian-critical-point}) turns out to
be equivalent to

$$ -\triangle
\bar{H}_{\alpha}+\bar{H}_{\alpha}\left(2\bar{H}_{\alpha}^2+2\bar{K}_{\alpha}
-4\bar{\mathsf{R}}_{\alpha}+\textrm{Ric}(\bar{N}_{\alpha},\bar{N}_{\alpha})\right)=0.$$

However, $\bar{\mathsf{R}}_{\alpha}=0$ because it is a mixed sectional
curvature in the semi-Riemannian product, \cite{ON}, namely
$\left(\mathscr{R}^{+},\frac{1}{f_{+}^2}g\right) =
\left(\tilde{\mathscr{R}}^{+},\frac{1}{f_{+}^2}g\right)\times
(\mathrm{H}^{+},dt^2)$. Also,
$\Sigma_{\alpha}=\left(trace(\alpha),\frac{1}{f_{+}^2}g\right)\times
(\mathrm{H}^{+},dt^2)$ is  a semi-Riemannian product and so
$\bar{K}_{\alpha}=0$. Recall  $d\Pi^{1}(\partial_s
\phi)=\alpha^{\prime}$, $d\Pi^{1}(\partial_t
\phi)=d\Pi^{2}(\bar{N}_{\alpha})=0$, and
$\left(\tilde{\mathscr{R}}^{+},\frac{1}{f_{+}^2}g\right)$ has
sectional curvature $1$, then we get
$\textrm{Ric}(\bar{N}_{\alpha},\bar{N}_{\alpha})=-1$. Bearing in
mind $d\Pi^{2}(\partial_s \phi)=0$, we compute the mean curvature:
$$\bar{\mathbf{H}}_{\alpha}=
\frac{1}{2}\,(\bar{\nabla}_{\partial_{s}\phi}\partial_{s}\phi)^{\bot}+\frac{1}{2}\,(\bar{\nabla}_{\partial_{t}\phi}\partial_{t}\phi)^{\bot}=
\frac{1}{2}((d\Pi^{1})^{-1}(\tilde{\nabla}_{\alpha^{\prime}}\alpha^{\prime}))^{\bot}=-\frac{1}{2}\,\tilde{k}_{\alpha}\bar{N}_{\alpha},$$
where $\bar{\nabla}$ is the Levi-Civita connection of
$\left(\mathscr{R}^{+},\frac{1}{f_{+}^2}g\right)$, $\tilde{\nabla}$
is the Levi-Civita connection of
$\left(\tilde{\mathscr{R}}^{+},\frac{1}{f_{+}^2}g\right)$ and
$\tilde{k}_{\alpha}$ is the curvature of $\alpha$ in
$\left(\tilde{\mathscr{R}}^{+},\frac{1}{f_{+}^2}g\right)$. As a
consequence, we obtain
$\bar{H}_{\alpha}=-\frac{1}{2}\,\tilde{k}_{\alpha}$ and
$\triangle\bar{H}_{\alpha}=-\frac{1}{2}\,\tilde{k}_{\alpha}^{\prime\prime}$.
By using all these computations, we obtain that $\Sigma_{\alpha}$ is
a solution of
$(\mathbf{I}_{\partial\Sigma_{\alpha}}^{-1}([a_1,a_2]\times\mathds{R},
\mathscr{R}^{+});\overline{\mathfrak{W}})$ if and only if
\begin{equation}\label{elastica-R}
2\tilde{k}_{\alpha}^{\prime\prime}-\tilde{k}_{\alpha}^3+2\tilde{k}_{\alpha}=0.
\end{equation}
Finally, we check that this equation characterizes the elasticae in
$\left(\tilde{\mathscr{R}}^{+},\frac{1}{f_{+}^2}g\right)$.

In Section \ref{Auno-solutiones}, we computed the equation of both
space-like and time-like elastic curves in
$(\mathbf{AdS}_2,\frac{1}{h^2}\,g)$, see
(\ref{elastica-Euler-Lagrange}). Recall that in that section, the
axis $\eje$ was time-like,
$\mathcal{B}=\{\vec{x},\vec{y},\vec{z}\}$ was an orthonormal
basis,
$\mathbf{AdS_2}=\{\vec{v}\in\mathds{L}^3\,:\,\langle\vec{v},\vec{y}\rangle>0,
 \, \langle\vec{v},\vec{z}\rangle=0\}$ and
 $h(p)=\langle\vec{p},\vec{y}\rangle$, being $\vec{p}$ the position vector of the point $p$.
 Now, notice that
 $\left(\tilde{\mathscr{R}}^{+},\hat{g}=-\frac{1}{f_{+}^2}g\right)$ is an
 anti de Sitter space, and, with this new metric, $\eje$ is also a
 time-like axis. So, a time-like curve in $\left(\tilde{\mathscr{R}}^{+},\hat{g}\right)$
 is an elastica if and only if
 $2\hat{k}^{\prime\prime}-\hat{k}^3+2\hat{k}=0$,
 where $\hat{k}$ is the geodesic curvature of the curve.
 Both $\hat{g}$ and $\frac{1}{f_{+}^2}g$ have the same Levi-Civita connection. Thus, by
 comparing the Frenet equations, it is easy to see that
 $\hat{k}=-\tilde{k}$, where $\tilde{k}$ is the geodesic curvature in $\left(\tilde{\mathscr{R}}^{+},\frac{1}{f_{+}^2}g\right)$.
 And so, a space-like curve is a critical point of $\int_{\alpha}\tilde{k}^2$ in
 $\left(\tilde{\mathscr{R}}^{+},\frac{1}{f_{+}^2}g\right)$, if and only
 if $$2\tilde{k}^{\prime\prime}-\tilde{k}^3+2\tilde{k}=0.$$
This ends the proof of the theorem. \end{pf}

\vspace{1em}

\noindent \textbf{Remark.} The solutions of
(\ref{elastica-R}) are $$
\bar{\kappa}_{\alpha}(s)=C\,\mathbf{cn}\left(\sqrt{1
    -\frac{C^2}{2}}\,(s-a_o),\tilde{C}\right),$$
where $C\in\mathds{R}\setminus\{-\sqrt{2},\sqrt{2}\}$ and $a_o\in\mathds{R}$
are arbitrary constants and $\tilde{C}^2=\frac{C^2}{2C^2 -4}$. If
$C^2<2$, $s\in\mathds{R}$; in other case, $s\in\mathds{R}\setminus
\displaystyle\bigcup_{n\in\mathds{Z}}\, \left\{
    a_o+\frac{(2n+1)\mathsf{E}^{\prime}}{\sqrt{\frac{C^2}{2}-1}}\,\,\right\}$, being $\textsf{E}^{\prime}$ the complete elliptic integral
     of first kind with modulus $\sqrt{1-\tilde{C}^2}$.

\section{$\mathbf{A}_2$-invariant Lorentzian solutions} \label{Ados-L-solutiones}

Along this section, we describe the class of
$\mathbf{A}_2$-invariant Lorentzian solutions. Firstly, we consider
the case when the surfaces are contained in one fundamental
region, i.~e., we deal with the fundamental solutions. Certainly, this
is the easy but basic case. As one can guess, two classification
theorems are obtained according to the nature of the profile
curve. On one hand, we obtain Lorentzian solutions,
$\Sigma_{\alpha}$, contained in either $\mathscr{R}^{+}$ or
$\mathscr{R}^{-}$ which are generated by time-like free elasticae
in the de Sitter plane. On the other hand, we also obtain a second
class of Lorentzian solutions, that lie in either $\mathscr{Q}^{+}$
or $\mathscr{Q}^{-}$, coming from free elasticae in the hyperbolic
plane. To be precise, we consider the following problem: Let
$\gamma$ be a time-like curve in either $\tilde{\mathscr{R}}^+$ or
$\tilde{\mathscr{R}}^-$, or a curve in either
$\tilde{\mathscr{Q}}^+$ or $\tilde{\mathscr{Q}}^-$. \textit{What
does it have to satisfy in order to make $\Sigma_{\gamma}$ be a
solution}? or equivalently, \textit{What does $\gamma$ have to
satisfy in order to make $\Sigma_{\gamma}$ be a Willmore surface
with prescribed space-like Gauss map along the boundary}?

The answer to this problem is given in the next pair of
statements. We omit their proofs because they are quite similar to
that of Theorem \ref{Riemannian-solutions} with only technical
changes.

\begin{thm} \label{Lorentzian-Fundamental-solutions-1} Let $\alpha$ be a time-like curve immersed in either $\tilde{\mathscr{R}}^{+}$ or
$\tilde{\mathscr{R}}^{-}$. Then, $\Sigma_{\alpha}$ is a Lorentzian
solution of the $O(2,1)$ Nonlinear Sigma Model, that is, a
Lorentzian Willmore surface with prescribed Gauss map along the
common boundary in $(\mathds{L}^3,g)$, if and only if, $\alpha$ is
a free elastica in the de Sitter plane
$\left(\tilde{\mathscr{R}}^{+},\frac{1}{f_{+}^2}g\right)$ or in
the de Sitter plane
$\left(\tilde{\mathscr{R}}^{-},\frac{1}{f_{-}^2}g\right)$, 
respectively.
\end{thm}
 
\begin{thm} \label{Lorentzian-Fundamental-solutions-2} Let $\beta$ be a curve immersed in either $\tilde{\mathscr{Q}}^{+}$ or
$\tilde{\mathscr{Q}}^{-}$. Then, $\Sigma_{\beta}$ is a Lorentzian
solution of the $O(2,1)$ Nonlinear Sigma Model, that is, a
Lorentzian Willmore surface with prescribed Gauss map along the
common boundary in $(\mathds{L}^3,g)$, if and only if, $\beta$ is
a free elastica in the  hyperbolic plane
$\left(\tilde{\mathscr{Q}}^{+},\frac{1}{h_{+}^2}g\right)$ or in
the hyperbolic plane
$\left(\tilde{\mathscr{Q}}^{-},\frac{1}{h_{-}^2}g\right)$, 
respectively.
\end{thm}

\noindent Nevertheless, it seems convenient to give the curvature
functions of the above solution generatrices.

\vspace{2mm}

\noindent \textbf{Time-like free elastic curves in}
$\left(\tilde{\mathscr{R}}^{+},\frac{1}{f_{+}^2}g\right)$
\textbf{and}
$\left(\tilde{\mathscr{R}}^{-},\frac{1}{f_{-}^2}g\right)$ are
those time-like curves with curvature
\begin{equation}\label{time-like-elastica-curvature-R}
    \tilde{\kappa}(s)=C\,\mathbf{cn}\left(i\sqrt{1+\frac{C^2}{2}}\,(s-a_o),\tilde{C}\right),
    \,\,\,\textrm{for}\,\, s\in \mathds{R}\setminus\bigcup_{n\in\mathds{Z}}\, \left\{
    a_o+\frac{(2n+1)}{\sqrt{1+\frac{C^2}{2}}}\textsf{E}^{\prime}\,\,\right\},
\end{equation} where $C\in\mathds{R}$, $a_o\in\mathds{R}$  are arbitrary
constants, $\tilde{C}^2=\frac{C^2}{2 C^2 + 4}$ and
$\textsf{E}^{\prime}$ is the complete elliptic integral of first
kind with modulus $\sqrt{1-\tilde{C}^2}$.

\vspace{2mm}

\noindent \textbf{Free elastic curves in}
$\left(\tilde{\mathscr{Q}}^{+},\frac{1}{h_{+}^2}g\right)$
\textbf{and}
$\left(\tilde{\mathscr{Q}}^{-},\frac{1}{h_{-}^2}g\right)$ are
those curves with curvature
\begin{equation}\label{elastica-curvature-Q}
    \tilde{\kappa}_{\gamma}(s)=C\,\mathbf{cn}\left(\sqrt{\frac{C^2}{2}-1}\,(s-a_o),\tilde{C}\right),\end{equation}
    $$\textrm{for}\,\,\,  s\in\, \left\{\begin{array}{lr}\mathds{R}\setminus
    \displaystyle\bigcup_{n\in\mathds{Z}}\, \left\{
    a_o
    +\frac{2n+1}{\sqrt{1-\frac{C^2}{2}}}\textsf{E}^{\prime}\right\}
    & \mathrm{ if }\,\,\,\,\,\,C^2<2 \\
    \\
    \mathds{R} & \mathrm{ if }\,\,\,\,\,\, C^2>2
    \end{array}\right\},$$
where $C\in\mathds{R}\setminus\{\sqrt{2},\sqrt{2}\}$ and $a_o\in\mathds{R}$
are arbitrary constants, $\tilde{C}^2=\frac{C^2}{2C^2 - 4}$ and
$\textsf{E}^{\prime}$ is the complete elliptic integral of first
kind with modulus $\sqrt{1-\tilde{C}^2}$.

\vspace{4mm}

 Moreover, we already know the existence of a wide family
of $\mathbf{A}_2$-invariant Lorentzian surfaces which are not
contained in a unique fundamental region. In other words, they are
obtained by pasting fundamental rotational surfaces. 
Then, we investigate the existence of
$\mathbf{A}_2$-invariant Lorentzian solutions which can be obtained by the
gluing mechanism. The following theorem provides the connection
between both the variational approach and the gluing mechanism.

\begin{thm} \label{Lorentzian-not-fundamental-solutions}
Given a surface $M$, let
$\phi\in\mathbf{I}_{\Gamma}^{1}(M,\mathds{L}^3)$ be a
$\mathbf{A}_2$-invariant immersion  whose image intersects at
least two different fundamental regions. Then, $(M,\phi)$ is a
Lorentzian solution of the $O(2,1)$ Nonlinear Sigma Model, that is,
a Lorentzian Willmore surface with prescribed Gauss map along the
boundary in $(\mathds{L}^3,g)$, if and only if, $\phi(M)$ is a
$\mathbf{A}_2$-invariant connected piece (with boundary) of one of the
following surfaces:
\begin{itemize}
\item A Lorentzian plane orthogonal to the axis. \item A
one-sheet hyperboloid with arbitrary radius, centered at any
point $p\in\eje$ and with axis $p+\langle
\overrightarrow{z}\rangle$ (see Subsection \ref{examples-motivation}).
\end{itemize}
\end{thm}

\begin{pf}
Given a surface $M$, and a   $\mathbf{A}_2$-invariant Lorentzian
immersion $\phi\in\mathbf{I}_{\Gamma}^{1}(M,\mathds{L}^3)$, we
know $\phi(M)$ is generated by a curve $\beta$ in $\mathscr{Q}$
and a countable family of time-like curves $\alpha_i$ in
$\mathscr{R}$, that are in general position. It should be noticed
(see Theorems \ref{Lorentzian-Fundamental-solutions-1} and
\ref{Lorentzian-Fundamental-solutions-2}) that $\phi$ is a critical
point of
$\left(\mathbf{I}_{\Gamma}^{1}(M,\mathds{L}^3);\mathfrak{S}\right)$,
if and only if, the following five conditions hold:
    \begin{enumerate}
    \item $\mathrm{trace}(\beta)\cap\tilde{\mathscr{Q}}^{+}$ consists of free
    elastic curves in
    $\left(\tilde{\mathscr{Q}}^{+},\frac{1}{h_{+}^2}g\right)$,
\item $\mathrm{trace}(\beta)\cap\tilde{\mathscr{Q}}^{-}$ consists of free
    elastic curves in
    $\left(\tilde{\mathscr{Q}}^{-},\frac{1}{h_{-}^2}g\right)$,
    \item $\mathrm{trace}(\alpha_i)\cap\tilde{\mathscr{R}}^{+}$ consists of time-like free
    elastic curves in
    $\left(\tilde{\mathscr{R}}^{+},\frac{1}{f_{+}^2}g\right)$
    $\forall$ $i$,
\item $\mathrm{trace}(\alpha_i)\cap\tilde{\mathscr{R}}^{-}$ consists of time-like free
    elastic curves in
    $\left(\tilde{\mathscr{R}}^{-},\frac{1}{f_{-}^2}g\right)$
    $\forall$ $i$, and
    \item $\phi\vert_{\mathsf{K}}$ is a critical point of
$\left(\mathbf{I}_{\phi(\partial
\mathsf{K})}^{1}(\mathsf{K},\mathds{L}^3);\mathfrak{W}^{\mathsf{K}}\right)$
for any non-null polygon $\mathsf{K}\subseteq M$, such that both
$\mathsf{K}\bigcap(\mathscr{R}^{+}
    \cup\mathscr{R}^{-})$ and $\mathsf{K}\bigcap(\mathscr{Q}^{+}
    \cup\mathscr{Q}^{-})$ are not empty.
    \end{enumerate}
The last condition holds if and only if, for any non-null polygon
$\mathsf{K}$,
(\ref{Scalar-Willmore-surfaces-prescribed-Gauss-map}) is satisfied
for all $\mathbf{V}\in
\mathsf{T}_{\phi}\left(\mathbf{I}_{\Gamma}^{1}(\mathsf{K},\mathds{L}^3)\right)$.
Notice that $N_{\phi}(\mathsf{R}^\mathbf{V})=0$ in this case, because
$\mathds{L}^3$ is flat. Therefore, if $\phi$ satisfies the first
four conditions, the last one is satisfied too.

Now, we wish to control the curves $\alpha_i$ for each $i$. It is
known, for each $\alpha_i$, the existence of a function,
$f_{i}:]-\varepsilon,\varepsilon[\rightarrow\mathbb{R}$, such that
$\alpha_i(t)=(f_{i}(t),0,t)$ is a parametrization of $\alpha_i$, in a neighborhood of
$\mathrm{trace}(\alpha_i)\bigcap\eje$.
Even more, we know that $f_{i}'(0)=0$ (see proof of Theorem
\ref{Pegado}). The curvature function of $\alpha_i
:]0,\varepsilon[\longrightarrow\left(\tilde{\mathscr{R}}^{+},\frac{1}{f_{+}^2}g\right)$,
can be computed to be
 $$\kappa_i(t):=\frac{-f_i'(t)+t
f_i''(t)+(f_i'(t))^3}{(1-(f_i'(t))^2)^{3/2}}.$$ Since
$f_{i}'(0)=0$, we obtain that $\lim_{t\rightarrow
0}\kappa_{i}(t)=0$. On the other hand, the curvature of a
time-like free elastic curve in
$\left(\tilde{\mathscr{R}}^{+},\frac{1}{f_{+}^2}g\right)$ is given
by (\ref{time-like-elastica-curvature-R}). If we use the
properties of the elliptic cosinus of Jacobi, it is easy to check
that the module of (\ref{time-like-elastica-curvature-R}) is equal
or greater than $|C|$. So, condition $3$ holds if, and only if,
$\alpha_i$ is a geodesic of
$\left(\tilde{\mathscr{R}}^{+},\frac{1}{f_{+}^2}g\right)$. A
similar reasoning works for $\alpha_i
:]-\varepsilon,0[\longrightarrow\left(\tilde{\mathscr{R}}^{-},\frac{1}{f_{-}^2}g\right)$.
As a conclusion, we obtain that conditions $3$ and $4$ hold if and
only if $\alpha_i$ is a geodesic in both
$\left(\tilde{\mathscr{R}}^{+},\frac{1}{f_{+}^2}g\right)$ and
$\left(\tilde{\mathscr{R}}^{-},\frac{1}{f_{-}^2}g\right)$ for all
$i$.

Next, we work with $\beta$. For each piece of $\mathrm{trace}(\beta)\bigcap\tilde{\mathscr{Q}}^+$, there exists $s_o$ in
the closure of the domain of $\beta$, such that $p=\lim_{s\to s_o}\beta(s)\in\eje$ and $\exists i\in\mathcal{I}$ satisfying $\alpha_i=\alpha_{s_o}$
(see Definition \ref{general-position}).
Then,
there exists a function $f_\beta$, defined on
$]-\varepsilon,\varepsilon[$, such that
$\beta(t)=(f_{\beta}(t),t,0)$ provides a parametrization of
$\beta$. Again, we compute the curvature function of $\beta
:]0,\varepsilon[\longrightarrow\left(\tilde{\mathscr{Q}}^{+},\frac{1}{h_{+}^2}g\right)$,
obtaining  
$$\kappa(t):=\frac{f_{\beta}'(t)-t
f_{\beta}''(t)+(f_{\beta}'(t))^3}{(1+(f_{\beta}'(t))^2)^{3/2}}.$$
Bearing in mind that $f_{\beta}'(0)=0$, (see the proof of
Theorem \ref{Pegado}), we also obtain  $\lim_{t\rightarrow
0}\kappa(t)=0$. We compare this expression with the
curvature function of an elastica in
$\left(\tilde{\mathscr{Q}}^{+},\frac{1}{h_{+}^2}g\right)$, see
(\ref{elastica-curvature-Q}). If $C^2<2$, then the absolute value
of the curvature is equal or greater than $|C|$; otherwise, if $s$
denotes the arc-length parameter, $\kappa(s)$ is a periodic
function that takes the value $0$ at $s=a_o+(2n+1)\mathsf{E}/
\sqrt{\frac{C^2}{2}-1}$, and the value $C$ at
$s=a_o+4n\mathsf{E}/\sqrt{\frac{C^2}{2}-1}$, for all
$n\in\mathds{Z}$, where $\mathsf{E}$ is the complete elliptic
integral of first kind with modulus $\tilde{C}$. Now, we combine
the behavior of (\ref{elastica-curvature-Q}) with both
$\lim_{t\rightarrow 0}\kappa(t)=0$ and
$\lim_{t\rightarrow 0}\beta(t)\in\eje$, to conclude
that condition $1$ holds if, and only if, $\beta$ is a geodesic in
$\left(\tilde{\mathscr{Q}}^{+},\frac{1}{h_{+}^2}g\right)$. A
similar argument can be used to see that condition $2$ holds if
and only if $\beta$ is a geodesic in
$\left(\tilde{\mathscr{Q}}^{-},\frac{1}{h_{-}^2}g\right)$.

At this point, we take advantage on the knowledge one has on the
geodesics of both the hyperbolic plane and the de Sitter one.
Notice that we are regarding both surfaces as half-plane 
Poincar\'e models. We recall that the geodesics of the hyperbolic
plane are those curves whose trace is either a ray perpendicular
to the boundary or half a circle centered at the boundary. In the de Sitter plane, the time-like geodesics are those curves
whose trace is either a ray perpendicular to the boundary or half
of any of the connected components of a  time-like hyperbola
centered at any point of the boundary. Consequently, condition $1$
holds if and only if, $\beta$ can be reparametrized, in
$\tilde{\mathscr{Q}}^{+}$, as either:

\begin{itemize}
\item $\beta(t)=(A,t,0)$  for $t\in]0,b[\subseteq]0,+\infty[$, where $A\in \mathds{R}$, or

\item $\beta(t)=(A+\rho\cos(t),\rho\sin(t),0)$ for $t\in]0,a[\subseteq]0,\pi[$, where $A\in \mathds{R}$ and $\rho >0$.
\end{itemize}
Certainly, the same argument can be used for condition $2$, just
by replacing $\tilde{\mathscr{Q}}^{+}$ by
$\tilde{\mathscr{Q}}^{-}$; \hspace{2mm} $]0,\infty[ \quad
\textrm{by} \quad ]-\infty,0[$ \hspace{2mm} and \hspace{2mm}
$]0,\pi[ \quad \textrm{by} \quad ]-\pi,0[$. Since $\beta$ is
$\mathcal{C}^{\infty}$, we conclude that $\beta$ intersects $\eje$
at least once, even more, it is either
\begin{itemize}\item a segment perpendicular to $\eje$, or
\item a connected piece of a circle in $(\mathscr{Q},g)$ centered at
any point of $\eje$.
\end{itemize}
As an important consequence, the cardinal of $\{\alpha_i:i\in
\mathcal{I}\}$ is either $1$ or $2$. 
Based on the geodesics of the de Sitter plane, a similar reasoning can be applied to each $\alpha_i$,  obtaining that conditions $3$ and $4$ hold if
and only if for each $i$, $\alpha_i$ not only intersects $\eje$
but also it is either:
\begin{itemize}
\item a connected segment in $(\mathscr{R},g)$ perpendicular to
$\eje$, or \item a connected piece of a connected component of a
time-like Lorentzian circle in $(\mathscr{R},g)$, centered at any
point of $\eje$.
\end{itemize}
The only remaining detail consists of finding which combinations of
parametrizations of $\beta$ and $\{\alpha_i:i\in \mathcal{I}\}$ give rise to $\mathcal{C}^{\infty}$ gluing functions (recall Theorem \ref{Pegado}). The proof of the following assertions are left to the reader:
\begin{enumerate}
\item If $\beta(t)=(A,t,0)$ for an arbitrary constant
$A\in\mathds{R}$, then the only $\alpha$ that gives a
$\mathcal{C}^{\infty}$ gluing function is $\alpha(t)=(A,0,t)$. In
this case, the surface generated by $\alpha$ and $\beta$ is a
$\mathbf{A}_2$-invariant connected piece of plane
$\{(A,y,z)/y,z\in\mathds{R}\}$. \item If $\beta$ is a piece of the
circle in $(\mathscr{Q},g)$ with radius $\rho>0$ and center
$(A,0,0)$, that contains the point $(A+\rho,0,0)$, then to obtain
a $\mathcal{C}^{\infty}$ gluing function in that point, $\alpha$
must be a piece of the future-pointing connected component of the
time-like Lorentzian circle with radius $\rho$ and center
$(A,0,0)$. \item If $\beta$ is a piece of the circle in
$(\mathscr{Q},g)$ with radius $\rho>0$ and center $(A,0,0)$, that
contains the point $(A-\rho,0,0)$, then, to obtain a
$\mathcal{C}^{\infty}$ gluing function in that point, $\alpha$
must be a piece of the past-pointing connected component of the
time-like Lorentzian circle with radius $\rho$ and center
$(A,0,0)$.
\end{enumerate}
Finally, in the last two cases, i.~e., when $\beta$ is a piece of
circle in $(\mathscr{Q},g)$, the surface generated by
$\{\beta\}\cup\{\alpha_i :i\in \mathcal{I}\}$ is just a
$\mathbf{A}_2$-invariant connected piece of a one-sheet hyperboloid with center
$p\in\eje$, and axis $p+\langle \overrightarrow{z}\rangle$.
\end{pf}

\begin{center}
\begin{tabular}{|c|c|c|} \hline
 & &\\
\includegraphics[scale=0.45]{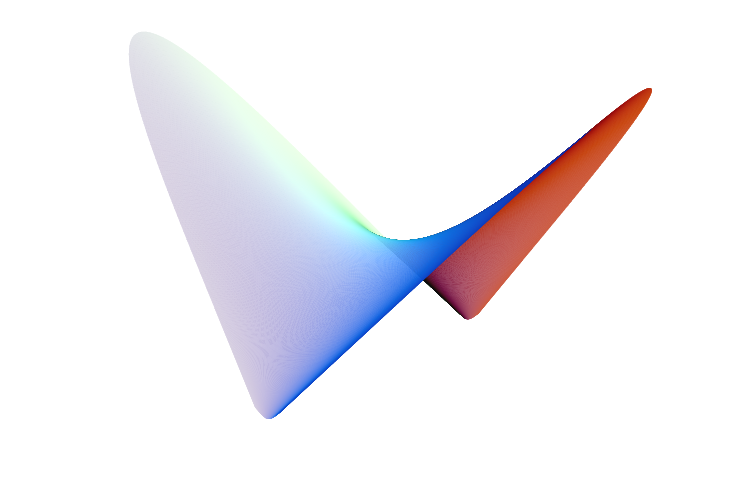} & \includegraphics[scale=0.25]{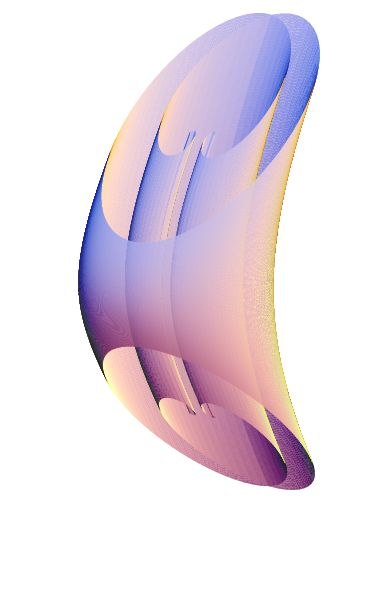}
& \includegraphics[scale=0.45]{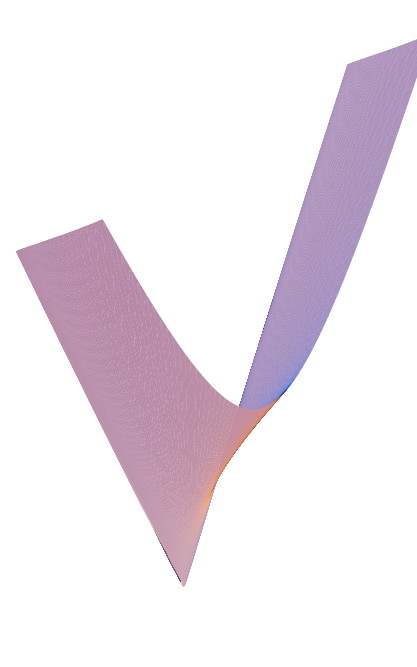}
\\
\hline
$\mathbf{A}_{2}$-invariant & $\mathbf{A}_{2}$-invariant Lorentzian  & $\mathbf{A}_{2}$-invariant Lorentzian \\
Riemannian solution & solution immersed in $\mathscr{Q}^+$ & solution immersed in $\mathscr{R}^+$\\
\hline
\end{tabular}
\end{center}

\section{$\mathbf{A}_{3}$-invariant Surfaces in $\mathds{L}^3$} \label{superficiesAtres}

In this section, we obtain, up to similarities in the
Lorentz-Minkowski space, the whole class of solutions of the
$O(2,1)$ Nonlinear Sigma Model which are symmetric under the group
$\mathbf{A}_{3}$. As  above, the first step consists of finding
the symmetric points, i.~e., the immersions that are
$\mathbf{A}_{3}$-invariant. In this sense, we will consider the
corresponding fundamental regions where one can get fundamental
symmetric surfaces,  well-known in the literature as rotational
surfaces with light-like axis, see for example \cite{Hano-Nomizu}.
Then, we will pay attention to the, a priori, reasonable problem
of gluing two fundamental symmetric surfaces which are contained
in different fundamental regions. Nevertheless, we will see that a
gluing mechanism does not work in this case, so we cannot paste
two rotational surfaces with light-like axis, lying in different
fundamental regions, to provide an $\mathbf{A}_{3}$-invariant
surface.

Given $\vec{x}\in\mathds{L}^3$ a light-like vector, we choose  a basis $\mathcal{B}=\{\vec{x},\vec{y},\vec{z}\}$,
such that: (1) $\vec{y}$ is a light-like vector
    with $\langle \vec{x},\vec{y}\rangle=-1$, and (2) $\vec{z}$ is
    a unitary space-like vector orthogonal to the plane
    $\textrm{Span}\{\vec{x},\vec{y}\}$. From now on, we will use coordinates with respect to
$\mathcal{B}$, so $g\equiv -2dx\,dy+dz^2$.  In $\mathds{L}^3\setminus\eje$ we will distinguish the
following  \textit{fundamental regions}:
$$\mathscr{S}^{+} = \{(x,y,z)\in\mathds{L}^3\, : \, y>0\}, \quad \mathrm{and} \quad  \mathscr{S}^{-} = \{(x,y,z)\in\mathds{L}^3\, : \, y<0\}.$$
Put $\mathscr{S}=\{(x,y,z)\in\mathds{L}^3\, : \, z=0\}$ and
$\mathscr{T}=\{(x,y,z)\in\mathds{L}^3\, : \, y=0\}$, and consider
the following open half planes:
\begin{eqnarray*}
  \tilde{\mathscr{S}}^{+} &=& \mathscr{S}^{+}\cap\mathscr{S}=\{(x,y,0)\in\mathds{L}^3\, : \, y>0\}, \,\, \mathrm{and}\\
  \tilde{\mathscr{S}}^{-} &=& \mathscr{S}^{-}\cap\mathscr{S}=\{(x,y,0)\in\mathds{L}^3\, : \,
  y<0\}.
    \end{eqnarray*}
  We also have the following parabolas, that are orbits under the
  action of $\mathbf{A}_{3}$:
  \begin{eqnarray*}
    \mathscr{P}^{+} &=& \{(x,1,z)\in\mathds{L}^3\, : \, -2x+z^2=0\}=\{(t^2 /2,1,t)\in\mathds{L}^3\, : \, t\in\mathds{R}\}, \,\, \mathrm{and} \\
    \mathscr{P}^{-} &=& \{(x,-1,z)\in\mathds{L}^3\, : \, 2x+z^2=0\}=\{(-t^2 /2,-1,t)\in\mathds{L}^3\, : \,
    t\in\mathds{R}\}.
  \end{eqnarray*}
  It should be noticed that $\mathscr{P}^{+}$ and
  $\mathscr{P}^{-}$ are space-like in $\mathds{L}^3$ with metric $dt^2$. Next, we define positive functions
  \begin{eqnarray*}
    l_{+}:\tilde{\mathscr{S}}^{+}\longrightarrow\mathds{R}, \quad l_{+}(x,y,0) &=& y, \,\,\mathrm{and} \\
    l_{-}:\tilde{\mathscr{S}}^{-}\longrightarrow\mathds{R}, \quad l_{-}(x,y,0) &=&
    -y.
    \end{eqnarray*}
  In this setting, it is not difficult to check  the following warped product decompositions
  \begin{eqnarray*}
    (\mathscr{S}^{+},g) &=& (\tilde{\mathscr{S}}^{+},g)\times_{l_{+}}(\mathscr{P}^{+},dt^2), \,\,\mathrm{and}\\
    (\mathscr{S}^{-},g) &=&
    (\tilde{\mathscr{S}}^{-},g)\times_{l_{-}}(\mathscr{P}^{-},dt^2).
  \end{eqnarray*}
  Furthermore, when we make the obvious conformal changes, it is
  easy to see that the surfaces $\left(\tilde{\mathscr{S}}^{+},\frac{1}{l_{+}^2}g\right)$
    and $\left(\tilde{\mathscr{S}}^{-},\frac{1}{l_{-}^2}g\right)$ are anti de Sitter planes with curvature $-1$.
Consequently, we obtain the following result.
  \begin{lem} \label{warped-product-null-axis}  \textit{$\left(\mathscr{S}^{+},\frac{1}{l_{+}^2}g\right)$ and
  $\left(\mathscr{S}^{-},\frac{1}{l_{-}^2}g\right)$ are
    semi-Riemannian products of an anti de Sitter plane and a space-like parabola.}
\end{lem}

\noindent The following result classifies the
$\mathbf{A}_{3}$-invariant surfaces. In particular, it proves that
the surfaces of this class lie in one fundamental region.

\begin{thm} \label{G3-Invariantes}
 Let $M$ be a connected surface and $\phi:M\longrightarrow\mathds{L}^3$ a non-degenerate immersion. Then,
 $(M,\phi^{*}(g))$ is $\mathbf{A}_3$-invariant if and only if one of the following statements hold:
\begin{enumerate}
    \item If $(M,\phi^{*}(g))$ is Riemannian, there exists a space-like curve, $\alpha$, immersed in either $\tilde{\mathscr{S}}^+$ or $\tilde{\mathscr{S}}^-$, such that
    $\phi(M)=\{\varsigma_t(\mathrm{trace}(\alpha))/t\in\mathds{R}\}$.
    \item If $(M,\phi^{*}(g))$ is Lorentzian, there exists a time-like curve, $\alpha$, immersed in either $\tilde{\mathscr{S}}^+$ or $\tilde{\mathscr{S}}^-$, such that
    $\phi(M)=\{\varsigma_t(\mathrm{trace}(\alpha))/t\in\mathds{R}\}$. \end{enumerate}
\end{thm}
\begin{pf} It is clear that, given any non-null curve, $\alpha$, immersed in either
$\tilde{\mathscr{S}}^{+}$ or in $\tilde{\mathscr{S}}^{-}$, then,
the surface parametrized by $X(s,t)=\varsigma_t(\alpha(s))$
provides an $\mathbf{A}_{3}$-invariant surface, \cite{Hano-Nomizu}.
If $\alpha$ is space-like, then the surface is Riemannian, and, if
$\alpha$ is time-like, the surface is Lorentzian. It is also easy
to see the converse, given an $\mathbf{A}_{3}$-invariant surface
immersed in either $\mathscr{S}^+$ or $\mathscr{S}^{-}$, then,
there exists a non-null curve, $\alpha$ immersed in either
$\tilde{\mathscr{S}}^{+}$ or $\tilde{\mathscr{S}}^{-}$ such that
the surface is parametrized as $X(s,t)=\varsigma_t(\alpha(s))$ and
so it is $\mathbf{A}_{3}$-invariant.

Thus, we only need to check that there do not exist
$\mathbf{A}_{3}$-invariant surfaces intersecting the plane
$\mathscr{T}$. First, notice that the orbits contained in
$\mathscr{T}$ are always light-like, so Riemannian
$\mathbf{A}_{3}$-invariant surfaces intersecting the plane
$\mathscr{T}$ cannot exist. 

On the other hand, let us assume there exists a Lorentzian
$\mathbf{A}_{3}$-invariant surface immersed in $\mathds{L}^3$,
$\phi(M)$, intersecting the plane $\mathscr{T}$. Then,
$\phi(M)\cap\mathscr{S}^{+}$ and $\phi(M)\cap\mathscr{S}^{-}$ are
the union of a countable family of $\mathbf{A}_{3}$-invariant
surfaces generated by time-like curves in
$\tilde{\mathscr{S}}^{+}$ and $\tilde{\mathscr{S}}^{-}$,
respectively. Note that the boundary of each of these curves has only
one point in $\eje$, because of its causality. Therefore, we must  check that none of the following cases hold:
 \begin{enumerate}
\item There exist two time-like curves $\alpha^{+}:]0,\delta[\longrightarrow\tilde{\mathscr{S}}^{+}$ and
$\alpha^{-}:]-\delta,0[\longrightarrow\tilde{\mathscr{S}}^{-}$ $(\delta>0)$ such that
\begin{enumerate}
\item $\lim_{s\rightarrow
0}\alpha^{+}(s)=\lim_{s\rightarrow 0}\alpha^{-}(s)$ is a point of
$\eje$. \item The surfaces generated by both curves can be glued,
obtaining a smooth $\mathbf{A}_{3}$-invariant Lorentzian  surface.
\end{enumerate}
\item There exist two time-like curves both in either
$\tilde{\mathscr{S}}^{+}$ or $\tilde{\mathscr{S}}^{-}$, satisfying
(a) and (b). \item There exists a time-like curve in either
$\tilde{\mathscr{S}}^{+}$ or $\tilde{\mathscr{S}}^{-}$, that
generates an $\mathbf{A}_{3}$-invariant surface that glue smoothly
with $\eje$, or a part of it, and the induced metric along the
union is Lorentzian.
\end{enumerate}
To do so, we study the behavior of the surfaces generated by a curve immersed in $\tilde{\mathscr{S}}^+$
and a curve immersed in
$\tilde{\mathscr{S}}^-$, in a neighborhood of $\eje$.
Let $\alpha^{+}:]0,\delta[\longrightarrow\tilde{\mathscr{S}}^{+}$ be a
time-like curve such that $\lim_{s\rightarrow
0}\alpha^{+}(s)\in\langle\vec{x}\rangle$. We define  $r^{+}$ as
$$r^{+}=\{\lim_{s\rightarrow 0}\varsigma_t(\alpha^{+}(s))/t\in\mathds{R}\}=\{\lim_{s\rightarrow
0}\alpha^{+}(s)+\lambda\vec{x}/\lambda>0\}.$$
Let $\alpha^{-}:]-\delta,0[\longrightarrow\tilde{\mathscr{S}}^{-}$ be a
time-like curve such that $\lim_{s\rightarrow
0}\alpha^{-}(s)\in\langle\vec{x}\rangle$.  We define  $r^{-}$ as
$$r^{-}=\{\lim_{s\rightarrow
0}\varsigma_t(\alpha^{-}(s))/t\in\mathds{R}\}=\{\lim_{s\rightarrow
0}\alpha^{-}(s)+\lambda\vec{x}/\lambda<0\}.$$
Then, cases $1$ and $2$ are not possible, because in both cases, the surfaces obtained after the gluing
are not a $\mathcal{C}^{\infty}$ surface in a
neighborhood of the point in which both curves glue.
A necessary condition for case $3$ to hold, is that the curve obtained
by gluing $\alpha$ and $r^+$ or $r^-$ (depending on if $\alpha$ is
immersed in $\tilde{\mathscr{S}}^+$ or $\tilde{\mathscr{S}}^-$,
respectively) must be $\mathcal{C}^{\infty}$. But in that case, the
tangent plane of the surface along $r^+$ or $r^-$ is $\{(x,0,z)/x,z\in\mathds{R}\}$, so the induced
metric is not Lorentzian along the union. \end{pf}

Finally, the last result classifies the $\mathbf{A}_3$-invariant
surfaces which are also solutions of the $O(2,1)$ Nonlinear Sigma Model.

\begin{thm} \label{G3-solutions} Let $\alpha$ be a non-null curve immersed in $\mathscr{S}$.  Then, $\Sigma_\alpha
=\{\varsigma_t(\mathrm{trace}(\alpha))/ t\in \mathds{R}\}$ is a
non-degenerate solution of the $O(2,1)$ Nonlinear Sigma Model, that
is, a non-degenerate Willmore surface with prescribed Gauss map
along the common boundary in $(\mathds{L}^3,g)$, if and only if
$\alpha$ is a free elastica in the anti de Sitter plane
$\left(\tilde{\mathscr{S}}^{+},\frac{1}{l_{+}^2}g\right)$
or in the anti de Sitter plane $\left(\tilde{\mathscr{S}}^{-},\frac{1}{l_{-}^2}g\right)$, respectively. 
\end{thm}

The proof is analogous to the one of Theorem \ref{Riemannian-solutions}, so it is left to the reader. We only have to recall
that the curvature function for free elasticae in the de Sitter plane were made explicit at the end of Section \ref{Auno-solutiones},
but in our case, we have to substitute  $\varepsilon_2$ by $\varepsilon$.

\begin{center}
\begin{tabular}{|c|c|} \hline
&\\
\includegraphics[scale=.25]{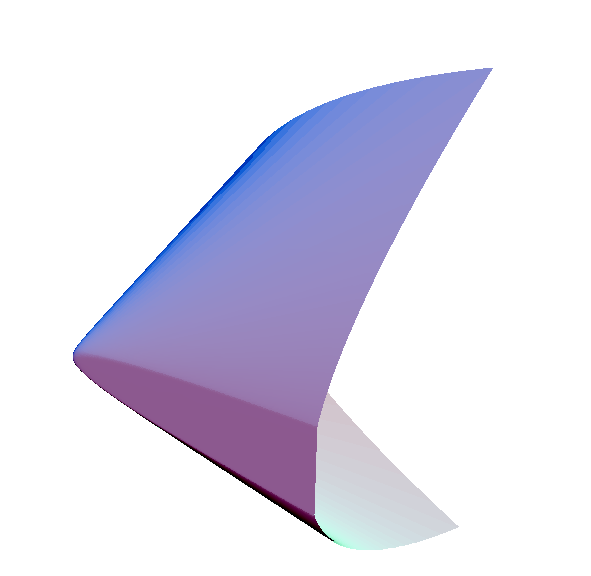}
  & \includegraphics[scale=.5]{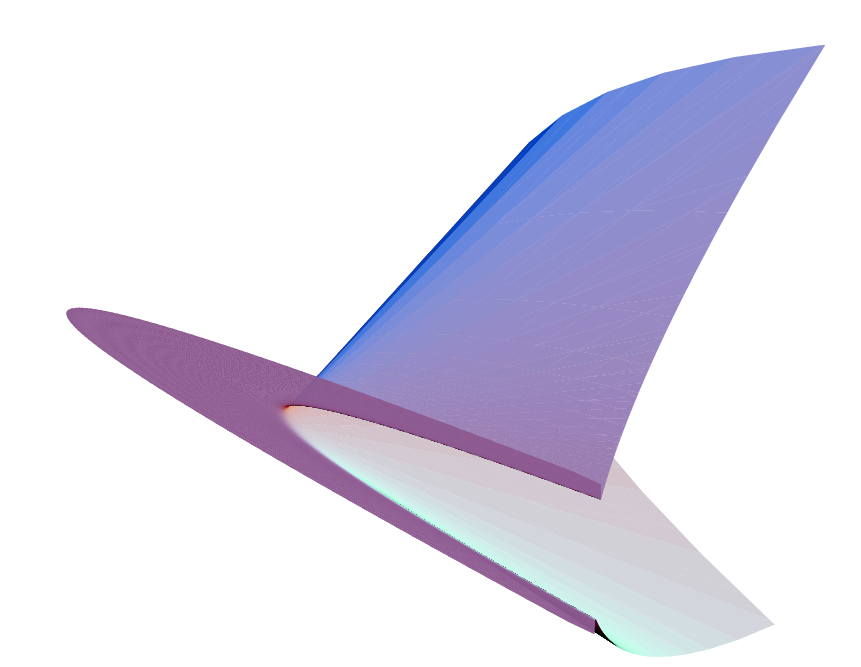}
 \\
&\\
\hline $\mathbf{A}_{3}$-invariant Riemannian solution & $\mathbf{A}_{3}$-invariant Lorentzian solution \\
\hline
\end{tabular}

\end{center}

\section{Appendix: A Gauss-Bonnet formula for non-null polygons} \label{apendice}

The Lorentzi\-an version of the Gauss-Bonnet
formula given in \cite{Birman-Nomizu} is only true when the
boundary, $\Gamma=\{\gamma_1,\gamma_2,\dots,\gamma_n\}$, is made
exclusively of time-like pieces. However, we are considering a
more general setting where some pieces of the boundary might be
space-like while others can be time-like. Therefore, we need to
extend the Gauss-Bonnet formula to this more general context.

Let $(S,\mathsf{g})$ be a Lorentzian surface. We choose an
orientation joint a time-orientation in $S$. For any unitary vector
$\vec{w}\in T_p S$, denote by $\vec{w}^{\perp}\in T_p S$ the
unique unitary vector such that
$\langle\vec{w},\vec{w}^{\perp}\rangle=0$ and the ordered basis
$\{\vec{w},\vec{w}^{\perp}\}$ is positively oriented. By choosing
such basis and expressing vectors by their corresponding
coordinates, we can define the concept of hyperbolic angle,
\cite{Birman-Nomizu}, made by a pair of time-like vectors. Let
$\vec{u},\vec{v}$ be two unitary time-like vectors, if they are
future-pointing (or past-pointing), the angle,
$\angle[\vec{u},\vec{v}]$, from $\vec{u}$ to $\vec{v}$ is the
number $\theta$ such that
\[A_{\theta}\cdot\vec{u}=\vec{v}, \quad \textrm{with} \quad A_{\theta}=\left(%
\begin{array}{cc}
  \cosh{\theta} & \sinh{\theta} \\
  \sinh{\theta} & \cosh{\theta} \\
\end{array}%
\right).\]

Once we have defined the angle between two future-pointing (or
past-pointing) unitary time-like vectors, we can define the angle
between arbitrary unitary vectors according to the following cases:
\cite{Nesovic-Petrovic-Verstraelen}:
\begin{enumerate}
    \item If $\vec{u}$ is future-pointing and $\vec{v}$
    past-pointing (or viceversa) unitary time-like vectors, define
    \[ \angle[\vec{u},\vec{v}]=\angle[\vec{u},-\vec{v}].\]
    \item If $\vec{u}$ and $\vec{v}$ are unitary space-like vectors,
    then $\vec{u}^{\perp}$ and $\vec{v}^{\perp}$ are unitary 
    time-like vectors and so, we define
    \[\angle[\vec{u},\vec{v}]=\angle[\vec{u}^{\perp},\vec{v}^{\perp}].\]
    \item Finally, if $\vec{u}$ is time-like and $\vec{v}$
    space-like, we define
    \[\angle[\vec{u},\vec{v}]=\angle[\vec{u},\vec{v}^{\perp}]; \quad \angle[\vec{v},\vec{u}]=\angle[\vec{v}^{\perp},\vec{u}].\]
\end{enumerate}

With these definitions, Lemma 1 of \cite{Birman-Nomizu} still
holds, see \cite{Nesovic-Petrovic-Verstraelen}.

The main purpose of the next step is to realize, \textit{\`{a} la
Euler}, the geodesic curvature of any non-null curve, $\delta(s)$,
in $S$. Without loss of
generality, we may assume that $\delta(s)$ is arclength 
parametrized, with Frenet apparatus
$\{T(s)=\delta'(s),T^{\perp}(s)\}$,  curvature function $\kappa$,
and  Frenet equations
\[\nabla_{T}T=\varepsilon_{2}\kappa\,T^{\perp}, \quad \nabla_{T}T^{\perp}=-\varepsilon_{1}\kappa\,T,\]
where $\nabla$ stands for the Levi-Civita connection of
$(S,\mathsf{g})$, $\varepsilon_1=\mathsf{g}(T,T)$ and
$\varepsilon_2=\mathsf{g}(T^{\perp},T^{\perp})$. On the other
hand, let $Z(s)$ be a unitary  time-like vector field parallel along
$\delta(s)$. By choosing $Z(0)$ future-pointing, then it is so at
every point because parallel displacement preserves
time-orientation. Notice that $Z^{\perp}(s)$ is also parallel
along $\delta(s)$. Now, if we denote by
$\varphi(s)=\angle[T(s),Z(s)]$, the conclusion is
\begin{equation}\label{Euler}
    \varphi'(s)=-\kappa(s).
\end{equation}
To check this formula,  we will distinguish two cases:
\begin{enumerate}
    \item If $\delta(s)$ is time-like, the proof can be found in \cite{Birman-Nomizu}.
    \item If $\delta(s)$ is space-like, then, in the basis $\{Z(s),Z^{\perp}(s)\}$, we have
     \begin{eqnarray}
\label{space-like1} A_{-\varphi(s)}\cdot Z(s)&=& \pm T^{\perp}(s), \\
 \label{space-like2} A_{-\varphi(s)}\cdot Z^{\perp}(s) &=& \mp T(s),
\end{eqnarray}
    depending on whether $T^{\perp}$ is future-pointing or past-pointing,  respectively.
    Next, we differentiate (\ref{space-like1}) with respect to $s$, compare it with
    the Frenet equations and finally, combine with (\ref{space-like2}) to obtain
    (\ref{Euler}).
\end{enumerate}
After this point, we can follow step by step the proof of
\cite{Birman-Nomizu}, valid for time-like polygons (i.~e., non-null
polygons with time-like boundary) to obtain the following result.

\vspace{2mm}

\noindent \textbf{Gauss-Bonnet formula for non null polygons}: Let
$(S,\mathbf{g})$ be a Lorentzian surface and let
$\mathsf{K}\subset S$ be a non-null polygon such that
$\partial\mathsf{K}$ is a simple closed curve made up of a finite
number of smooth non-null curves, $\delta_{j}(s)$, $1\leq j\leq
r$. Suppose that $\delta_{j}$ starts at $p_j\in S$ with initial
unitary  speed $\vec{t}_j$ and ends at $p_{j+1}\in S$ with terminal
unitary  speed $\vec{u}_j$, where $p_{r+1}=p_1$. Denote the exterior
angles at vertices as follows:
$\theta_1=\angle[\vec{u}_1,\vec{t}_2]$,
$\theta_2=\angle[\vec{u}_2,\vec{t}_3]$,$\dots$,$\theta_{r-1}=\angle[\vec{u}_{r-1},\vec{t}_r]$
and $\theta_r=\angle[\vec{u}_r,\vec{t}_1]$. In this framework, we
have
$$
    -\int_{\mathsf{K}}\,K\,dA +\int_{\partial\mathsf{K}}\,\kappa \,ds+\sum_{j=1}^r\,\theta_j=0,
$$ where $K$ stands for the Gaussian curvature of $(S,\mathsf{g})$ and $\kappa$ denotes the geodesic curvature
 along $\partial\mathsf{K}$. 

\end{document}